\documentclass[prd,amsmath,superscriptaddress,twocolumn]{revtex4}

\usepackage{hyperref}
\usepackage{ifthen}
\usepackage{amsmath}
\usepackage{amssymb}
\usepackage{graphics}
\usepackage{color}

\definecolor{darkgreen}{cmyk}{0.85,0.2,1.00,0.2}
\definecolor{purple}{cmyk}{0.5,1.0,0,0}

\begin{document}

\title{Consistency check of $\Lambda$CDM phenomenology}
\author{Lucas Lombriser}
\affiliation{Institute for Theoretical Physics, University of Z\"{u}rich, Winterthurerstrasse 190, CH-8057 Z\"{u}rich, Switzerland}

\date{\today}

\begin{abstract}

The standard model of cosmology $\Lambda$CDM assumes general relativity, flat space, and the presence of a positive cosmological constant. We relax these assumptions allowing spatial curvature, a time-dependent effective dark energy equation of state, as well as modifications of the Poisson equation for the lensing potential, and modifications of the growth of linear matter density perturbations in alternate combinations. Using six parameters characterizing these relations, we check $\Lambda$CDM for consistency utilizing cosmic microwave background anisotropies, cross correlations thereof with high-redshift galaxies through the integrated Sachs-Wolfe effect, the Hubble constant, supernovae, and baryon acoustic oscillation distances, as well as the relation between weak gravitational lensing and galaxy flows. In all scenarios, we find consistency of the concordance model at the 95\% confidence level. However, we emphasize that constraining supplementary background parameters and parametrizations of the growth of large-scale structure separately may lead to \emph{a priori} exclusion of viable departures from the concordance model.

\end{abstract}

\maketitle

\section{Introduction}

The detection of the late-time acceleration of our Universe~\cite{riess:98, perlmutter:98} challenges the known laws of physics. General relativity and the standard model of particle physics cannot amount to the observed cosmic expansion unless we are willing to accept a seemingly random constant in the Einstein field equations or the presence of an unknown form of energy, along with large amounts of dark matter. The concordance model assumes such a cosmological constant or vacuum energy and provides a simple but successful description for the observed Universe. Given the lack of a complete understanding, it is important to repeatedly test this model against observations with the ambition of distinguishing between different explanations of the observed cosmic acceleration, e.g., a cosmological constant, dynamical dark energy, or a modification of gravity, and rule out or constrain the different models. See~\cite{tsujikawa:10, jain:10, sapone:10} for recent reviews on tests of nonstandard cosmologies.

Phenomenological parametrizations for departures from the concordance model have been studied in~\cite{chevallier:00, linder:02, linder:05, koivisto:05, caldwell:07, zhang:07, amendola:07, diporto:07, hu:07b, linder:09}. In this paper, we shall mainly adopt the parametrization and notation of~\cite{amendola:07, diporto:07}. In addition to the usual cosmological parameters, including spatial curvature, we use five phenomenological parameters quantifying modifications of the Poisson equation for the lensing potential, modifications of the growth of linear matter density perturbations, as well as a time-dependent effective dark energy equation of state. The parameters are based on models of modified gravity, but may also describe properties of dark energy and provide a framework to search for phenomena which may indicate new physical effects in current and future cosmological observations. 

We conduct a Markov chain Monte Carlo (MCMC) study of this parameter space using data from the cosmic microwave background (CMB) anisotropies, supernovae distances, the baryon acoustic oscillations (BAO) distances, and the Hubble constant. We also utilize information from the cross correlation between high-redshift galaxies and the CMB through the integrated Sachs-Wolfe (ISW) effect, as well as a probe of the relation between weak gravitational lensing and galaxy flows. For the predictions of the CMB anisotropies, we connect our phenomenological parameters to the parametrized post-Friedmann (PPF) framework~\cite{hu:07b, hu:08} and use its implementation into a standard Einstein-Boltzmann linear theory solver~\cite{fang:08b}.

In~\textsection\ref{sec:theory}, we define the phenomenological parametrization of the modifications to standard cosmology and explain their implications on various cosmological probes in~\textsection\ref{sec:predictions}. Modifications to the {\sc iswwll}~\cite{ho:08, hirata:08} code used for the galaxy-ISW (gISW) cross correlation observations are specified in~\textsection\ref{sec:gISW}. For the connection of our parametrization to the PPF formalism, we refer to the appendix. We present the results of our MCMC study in~\textsection\ref{sec:results} and discuss them in~\textsection\ref{sec:discussion}.

\section{Phenomenological modifications}\label{sec:theory}

We consider scalar linear perturbations of the Friedmann-Lema\^itre-Robertson-Walker background in the longitudinal gauge, hence
\begin{equation}
ds^2 = -(1+2\Psi)dt^2 + a^2(1+2\Phi)d\vec{x}^2,
\end{equation}
where $d\vec{x}^2$ is the unperturbed spatial line element with curvature $k$ and $c=1$.
Motivated by modified gravity models, effects from a nonstandard cosmology may introduce the following three time- and scale-dependent phenomenological modifications on the background and at quasistatic scales of linear perturbations (see, e.g.,~\cite{caldwell:07, zhang:07, amendola:07, hu:07b, koivisto:07, koivisto:08}; cf.~\cite{koivisto:05}):
\begin{itemize}
\item A deviation from the $\Lambda$CDM expansion history, parametrized by an effective dark energy equation of state $w_{\rm eff}(a)$,
\begin{eqnarray}
\left(\frac{H}{H_0}\right)^2 & = & \Omega_{\rm m} a^{-3} + \Omega_{\rm k} a^{-2} \nonumber \\
& &+ (1-\Omega_{\rm m}-\Omega_{\rm k}) a^{-3[w_{\rm eff}(a)+1]}.
\label{eq:expansionhistory}
\end{eqnarray}
\item An effective change of the Newton's constant, which we can parametrize by a free function $Q(k,a)$, yielding a generalized Poisson equation,
\begin{equation}
k^2 \Phi = 4\pi G a^2 Q \rho_{\rm m} \Delta_{\rm m}.
\label{eq:poisson_gen}
\end{equation}
\item A difference in the scalar linear potentials $\Psi$ and $\Phi$ parametrized by the free function $\eta(k,a)$,
\begin{equation}
\Psi = -(1+\eta) \Phi.
\label{eq:potentials_gen}
\end{equation}
\end{itemize}
Deviations from $Q=1$ and $\eta=0$, the standard values, do not necessary indicate modifications of general relativity but might, for instance, also originate from contributions of nonmatter fluids to the Poisson equation ($Q\neq1$), e.g., from clustering of dark energy or interactions between the dark components (see, e.g.,~\cite{song:10}),
or nonvanishing anisotropic stress ($\eta\neq0$) (see, e.g.,~\cite{koivisto:05, mota:07}). Dark energy models other than the cosmological constant also predict departures from $w_{\rm eff}(a)=-1$.
Note that a dark energy model can always be constructed to be formally equivalent to a modification of gravity through an effective dark energy stress-energy tensor. The parameters introduced here cannot distinguish between the two descriptions. Consideration of microphysical aspects may, however, indicate which is the more reasonable picture (see, e.g.,~\cite{hu:09, hu:07b}).

We can combine Eqs.~(\ref{eq:poisson_gen}) and (\ref{eq:potentials_gen}) to obtain~\cite{amendola:07}
\begin{equation}
k^2 \Phi_- = \frac{3H_0^2\Omega_{\rm m}}{2a} \Sigma \Delta_{\rm m},
\label{eq:Sigma_par}
\end{equation}
where $\Sigma = Q(1+\eta/2)$ and $\Phi_- = (\Phi-\Psi)/2$.
This relation is in particular sensitive to weak lensing measurements and the ISW effect can be used to probe its time evolution.
Modifications as in Eqs.~(\ref{eq:poisson_gen}) and (\ref{eq:potentials_gen}) lead to changes in the growth of the linear matter overdensity perturbation $\Delta_{\rm m}$, which we parametrize via the growth index $\gamma$~\cite{linder:05, diporto:07},
\begin{equation}
\frac{d \ln \Delta_{\rm m}}{d \ln a} = \Omega_{\rm m}(a)^{\gamma} (1+\xi),
\label{eq:gamma_par}
\end{equation}
where $\Omega_{\rm m}(a) = H_0^2\Omega_{\rm m} a^{-3} H^{-2}$. We also introduce here the parameter $\xi$ to account for growth rates beyond unity, given $\gamma>0$ and $0\leq\Omega_{\rm m}\leq1$, as can be observed in scalar-tensor theories~\cite{diporto:07}. Hence, detection of $\xi>0$ may indicate the presence of an additional attractive force.

We decide to use the quantities defined in Eqs.~(\ref{eq:Sigma_par}) and (\ref{eq:gamma_par}) to scan the cosmological data for departure from standard theory and replace the two free functions $Q$ and $\eta$ by $\Sigma$ and $(\gamma,\xi)$.
Note that $\xi$ should not be interpreted as
a new parameter in the gravitational dynamics like $Q$ and $\eta$, but rather as the separation $\gamma \rightarrow \gamma + \Gamma$, where we assume the specific form $\Gamma = \ln(1+\xi)/\ln\Omega_{\rm m}(a)$.

From the ordinary differential equation, which describes the correct behavior of the linear matter overdensity perturbations in the quasistatic regime,
\begin{equation}
\Delta_{\rm m}'' + \left( 2 + \frac{H'}{H} \right) \Delta_{\rm m}' - \frac{3}{2} \frac{H_0^2 \Omega_{\rm m}}{ a^3 H^2 } \mathcal{F} \Delta_{\rm m} = 0,
\end{equation}
where $\mathcal{F}= Q(1+\eta) = 2\Sigma (1+\eta)/(2+\eta)$ and primes denote derivatives with respect to $\ln a$, we derive
\begin{eqnarray}
\mathcal{F} & = & \frac{2}{3} \Omega_{\rm m}(a)^{2\gamma-1} (1+\xi)^2 - 
\frac{2}{3} \left\{ \gamma \left[2\frac{H'}{H}+3\right] \right. \nonumber\\
& & \left. - \left[\frac{H'}{H}+2\right] \right\} \Omega_{\rm m}(a)^{\gamma-1} (1+\xi),
\label{eq:Fcal}
\end{eqnarray}
where we assumed constant $\gamma$ and $\xi$. Given $\Sigma$ and $(\gamma$, $\xi)$, we can use Eq.~(\ref{eq:Fcal}) to derive $Q$ and $\eta$ and thus define our modifications to $\Lambda$CDM in the framework of Eqs.~(\ref{eq:expansionhistory}) through (\ref{eq:potentials_gen}). Note that
\begin{equation}
\mathcal{F} = \frac{2}{3}(1+\xi)^2 + \frac{1}{3} (1+\xi),
\label{eq:mathcalF_lim}
\end{equation}
whenever $a^{w_{\rm eff}(a)} \gg 1$ and $w_{\rm eff}'(\ln a) \ll 1$ as $a\ll1$. If $|\xi|\ll1$ at $a\ll1$, standard gravity is reproduced. Otherwise, modifications to gravity persist up to high redshifts.

\subsection{Parametrization}\label{sec:parametrization}

It is a difficult task to find general functions with a minimal set of free parameters that are flexible enough to capture the wealth of possible modifications in Eqs.~(\ref{eq:expansionhistory}) through (\ref{eq:potentials_gen}) or equivalently in Eqs.~(\ref{eq:expansionhistory}), (\ref{eq:Sigma_par}), and (\ref{eq:gamma_par}). Rather than to construct such a function for each relation, we decide to use a low number of five parameters, in addition to spatial curvature, and examine the combinational aspects of introducing modifications in each of the relations. We restrict to constant and time-dependent modifications and choose the parametrization in a way that $\Lambda$CDM is contained in the parameter space.

For the expansion history we consider the parametrization
\begin{eqnarray}
H(a)^2 & = & H_0^2 \Omega_{\rm m} a^{-3} + H_0^2 \Omega_{\rm k} a^{-2} + H_0^2 (1-\Omega_{\rm m}-\Omega_{\rm k}) \nonumber\\
& & \times a^{-3(1+w_0+w_a)} \exp{\left[ 3w_a (a-1) \right]},
\label{eq:hubbleexpansion}
\end{eqnarray}
where the dark energy equation of state is given by~\cite{chevallier:00, linder:02}
\begin{eqnarray}
w_{\rm DE}(a) & = & w_0 + (1-a) w_a, \\
a^{-3[1+w_{\rm eff}(a)]} & = & \exp \left[ 3 \int_a^1 \frac{1+w_{\rm DE}(a')}{a'} da' \right].
\end{eqnarray}
Note that this parametrization does not always provide a good fit to model predictions. We adopt the approach primarily due to its simplicity and wide usage. In the limit $(w_0,w_a)\rightarrow(-1,0)$, the effective dark energy term in Eq.~(\ref{eq:hubbleexpansion}) reduces to a cosmological constant. The parametrization is less successful, e.g., in the case of Dvali-Gabadadze-Porrati (DGP) gravity~\cite{dvali:00}. It provides a good approach for distance measures~\cite{linder:05} and the quasistatic regime for the self-accelerated branch, but it fails for the normal branch due to the appearance of a divergence in $w_{\rm DE}(a)$. Also note that choosing a specific form for $w_{\rm DE}$ may have a nontrivial effect on the constraints inferred for it (see, e.g.,~\cite{nesseris:10} for a model-independent approach).

Next, we need to define modifications to the rate of growth and the Poisson equation. We decide to use a constant growth index $\gamma_0$, which is a good approximation to general relativity, where $\gamma_0\approx0.55$, or the quasistatic regime of self-accelerating DGP gravity~\cite{deffayet:00}, where $\gamma_0\approx0.68$~\cite{lue:04, linder:05, linder:07}. We further use a constant $\xi=\xi_0$, which was found to provide a good fit to scalar-tensor theories where the scalar field couples to dark matter~\cite{diporto:07}.
Equation~(\ref{eq:gamma_par}) relates to the parametrization of~\cite{linder:09} as $d \ln g_{\star}/ d \ln a = \xi \Omega_m(a)^{\gamma}$, which was introduced to describe effects from early dark energy, modified gravity at high redshifts, or early acceleration. For these models, $g_{\star}$ was found to be well-described by a constant~\cite{linder:09}. Therefore, there is only limited correspondence to a constant $\xi=\xi_0=0$.
For $\Sigma$, we use a parametrization that reduces to its general relativistic value $\Sigma \rightarrow 1$ at early times.
Hence, for our consistency test, we furthermore set
\begin{eqnarray}
\gamma & = & \gamma_0, \\
\xi & = & \xi_0 \\
\Sigma & = & 1 + \Sigma_0 a.
\end{eqnarray}
For reference, we denote our nonstandard cosmologies by the extra parameters we allow to be free. Hence, when, e.g., taking $\Omega_{\rm k}$ and $\gamma_0$, as well as $w_0$ and $w_a$, to be free parameters deviating from their standard values, we denote the according model by $\gamma w {\rm k}$.
Note that in the limit where $\{w_0,\ w_a,\ \gamma_0,\ \xi_0,\ \Sigma_0) \rightarrow \{-1,\ 0,\ 0.55,\ 0,\ 0)$, Eqs.~(\ref{eq:expansionhistory}), (\ref{eq:Sigma_par}), and (\ref{eq:gamma_par}) reduce to general relativity with cold dark matter and a cosmological constant. From Eqs.~(\ref{eq:gamma_par}) and (\ref{eq:hubbleexpansion}), it is clear that when $w_a\approx-w_0$, there is a degeneracy between $w_a$ and $\xi_0$, which manifests itself, in particular, at high redshifts. To explore the parameter space unclosed by free $w_0$, $w_a$, and $\xi_0$, we furthermore study three models where we fix $\gamma_0=0.55$ and $\Sigma_0=0$ while allowing the following degrees of freedom:
\begin{itemize}
 \item[(A)] $\xi_0$, $w_{\rm DE} = w_0+(1-a)w_a$,
 \item[(B)] $\xi_0$, $w_{\rm DE} = -1+(1-a)w_a$,
 \item[(C)] $\xi_0$, $w_{\rm DE} = -1+\lambda_0(1-1.15a)$,
\end{itemize}
such that $\Lambda$CDM is a limiting case in all of the three models. The slope of $w_{\rm DE}(a)$ in model (C) is motivated by the best-fit values derived in~\textsection\ref{sec:results}. 
For numerical predictions, we connect our parametrizations to the linear PPF framework as described in Appendix~\ref{sec:ppf_connection}.

Note that we have ignored scale dependence of the parameters, which is a good approximation within $\Lambda$CDM. Modifications of gravity such as $f(R)$ gravity models~\cite{carroll:03, nojiri:03, capozziello:03}, however, may introduce a strong scale dependence in these relations. The same holds for DGP gravity, where the deficiency is, however, restricted to near-horizon and superhorizon scales~\cite{seahra:10}. Although for these cases the parametrization is not very descriptive, it still serves as a useful tracer of nonstandard phenomenologies. Departure from the standard parameter values may indicate inconsistencies in the concordance model and point toward new physical effects, which in turn have to be addressed with more developed theories. It has been pointed out, by conducting a principal component analysis~\cite{zhao:09}, that such inconsistencies in Eqs.~(\ref{eq:poisson_gen}) through (\ref{eq:Sigma_par}) are more likely to be detected in scale-dependent modifications due to weaker sensitivity in the data to time-dependent deviations~\cite{pogosian:10, zhao:10}. A parametrization like the one described in this section is still capable of tracing trends in the observations and disclosing new regions in the parameter space of possible modifications, especially when simultaneously allowing for time-dependent modifications in the effective dark energy equation of state.

\section{Consistency check}\label{sec:constraints}

We use a variety of cosmological data sets to check against nonstandard cosmology. First we use the CMB anisotropy data from the seven-year Wilkinson Microwave Anisotropy Probe (WMAP)~\cite{WMAP:10}, the Arcminute Cosmology Bolometer Array Receiver (ACBAR)~\cite{ACBAR:07}, the Balloon Observations Of Millimetric Extragalactic Radiation and Geophysics (BOOMERanG) flight in 2003 (B03)~\cite{B03:05}, and the Cosmic Background Imager (CBI)~\cite{CBI:04}. Next we employ data from the Supernova Cosmology Project (SCP) Union2~\cite{UNION2:10} compilation, the measurement of the Hubble constant from the Supernovae and $H_0$ for the Equation of State (SHOES)~\cite{SHOES:09} program generalized by~\cite{reid:09}, and the BAO distance measurements of~\cite{BAO:09}. Furthermore, we take gISW cross correlation observations using the {\sc iswwll} code of~\cite{ho:08, hirata:08}, and the $E_G$ measurement, probing the relation between weak gravitational lensing and galaxy flows, of~\cite{reyes:10}.

Note that we restrict to data sets amenable to linear perturbation theory. Nonlinear probes such as from the abundance of clusters and the full range of scales of weak gravitational lensing yield tight constraints on the parameters that quantify modifications to general relativity (see, e.g.,~\cite{rapetti:08, schmidt:09, rapetti:09, daniel:10a, daniel:10b, lombriser:10}). In the case of $f(R)$ gravity models or DGP gravity, nonlinear effects have been studied in~\cite{oyaizu:08a, oyaizu:08b, schmidt:08b, schmidt:09, schmidt:09b, schmidt:09c, koyama:09, beynon:09, ferraro:10, zhao:10b} and spherical collapse within various dark energy models in, e.g.,~\cite{pace:10, wintergerst:10}. However, for phenomenological parametrizations like the one we use, nonlinear behavior has not been examined in full extent. Hence, applying standard scaling relations may lead to illusive conclusions. Furthermore, note that the inclusion of gISW cross correlations, along with the ISW effect in the CMB, and $E_G$, yield competitive results to constraints from nonlinear probes (cf., e.g.,~\cite{bean:10}).

In~\textsection\ref{sec:predictions} we discuss the predictions for some of these observables for specific parameter values. In~\textsection\ref{sec:results} we present the results of a MCMC likelihood analysis, which is conducted with the publicly available {\sc cosmomc}~\cite{COSMOMC:02} package.

\subsection{Cosmological observables}\label{sec:predictions}

In this section we illustrate model predictions of various cosmological observables we use to derive our results. We chose the parameters of the various models that highlight results from the MCMC analysis.

As our basic set, we choose a parametrization that separates high-redshift and low-redshift constraints. Specifically we take six high-redshift parameters: the physical baryon and cold dark matter density $\Omega_bh^2$ and $\Omega_ch^2$, the ratio of sound horizon to angular diameter distance at recombination $\theta/100$, the optical depth to reionization $\tau$, the scalar tilt $n_s$, and amplitude $A_s$ at $k_* = 0.002~\textrm{Mpc}^{-1}$. We extend this parameter set with alternate combinations of the six additional parameters, describing departures from the concordance model: the spatial curvature density $\Omega_{\rm k}$, two parameters for the evolution of the effective dark energy equation of state, $w_0$ and $w_a$, the growth index $\gamma_0$ and the scaling of the growth rate $\xi_0$, as well as the first-order of a time-dependent modification of the Poisson equation for the lensing potential $\Sigma_0$.

We illustrate predictions from the maximum-likelihood $\Lambda$CDM model, as well as from the overall best-fit model (see~\textsection\ref{sec:results}). Hereby, we derive the parameter values using the full set of cosmological data. To demonstrate effects from the variation of a specific basic cosmological parameter on the observables, we use the corresponding 1D-marginalized 68\% confidence limits from the $\Lambda$CDM model while setting the complementary parameters to their best-fit values. For a supplementary parameter, we use its 1D-marginalized 68\% confidence boundary, when including it as the sole extra parameter while setting the basic parameters to their $\Lambda$CDM best-fit values. Note, however, that for the nonstandard cases, $\gamma_0$ is always simultaneously varied with any of the extra parameters. Hereby, $(w_0,w_a)$ should be counted as one parameter; i.e., $w_0$ and $w_a$ are always either fixed to $(-1,0)$ or both considered free.

\subsubsection{Distance measures}

\begin{figure}
 \resizebox{\hsize}{!}{\includegraphics{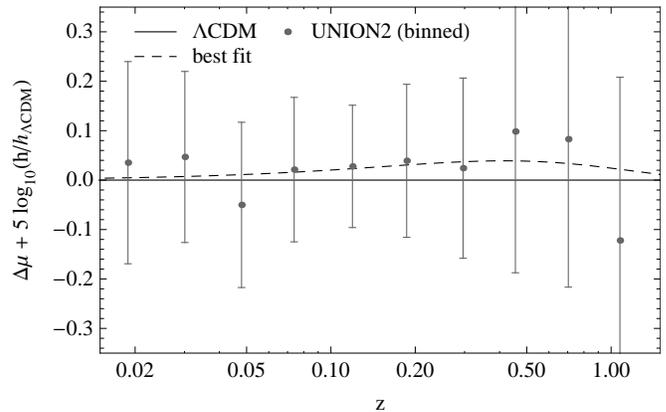}}
 \caption{Overall best-fit distance modulus with respect to the best-fit $\Lambda$CDM distance modulus. For illustration, the UNION2 data are binned into ten data bands with logarithmic spacing.}
 \label{fig:UNION2}
\end{figure}

\begin{figure}
 \resizebox{.75\hsize}{!}{\includegraphics{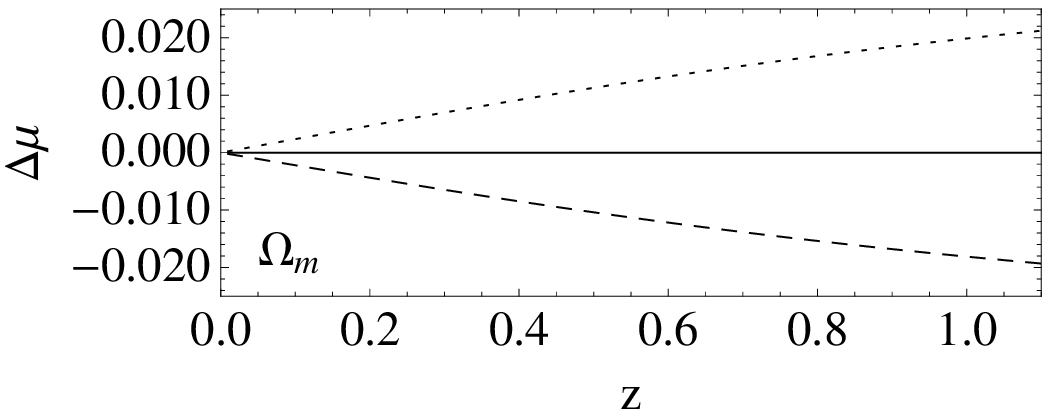}}
 \resizebox{.75\hsize}{!}{\includegraphics{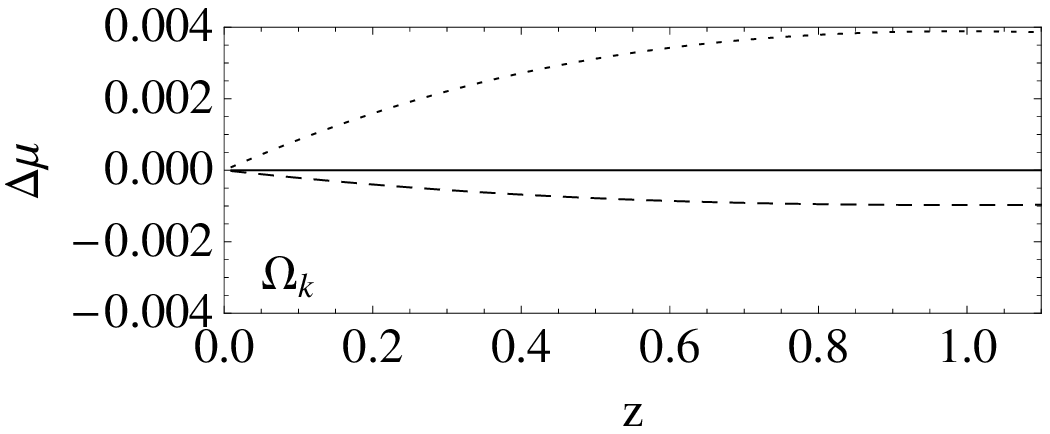}}
 \resizebox{.75\hsize}{!}{\includegraphics{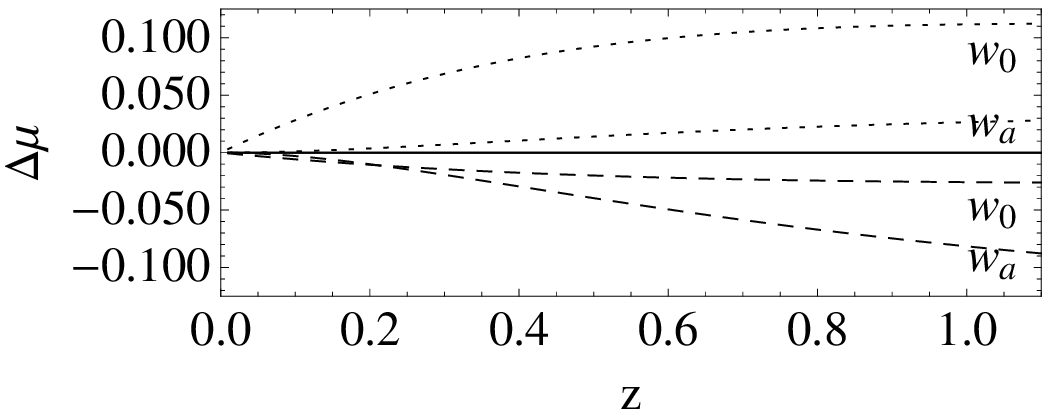}}
 \caption{Deviation of the distance modulus with respect to the overall best-fit $\Lambda$CDM model. The dashed and dotted lines indicate upper and lower 1D-marginalized 68\% confidence limits of the according parameters, respectively, while other parameters are fixed to their $\Lambda$CDM best-fit values.}
 \label{fig:UNION}
\end{figure}

The comparison of the magnitudes of high-redshift to low-redshift supernovae yields a relative distance measure. Theoretical predictions for the distance modulus are related to the luminosity distance, $d_L(z) = (1+z) r(z)$, where $r(z)$ is the comoving angular diameter distance defined by
\begin{equation}
r(z) = \left\{
\begin{array}{ll}
\sin\left[H_0 \sqrt{-\Omega_{\rm k}} \chi(z)\right]/H_0\sqrt{|\Omega_{\rm k}|}, & \Omega_{\rm k} < 0, \\
\chi(z), & \Omega_{\rm k} = 0, \\
\sinh\left[H_0 \sqrt{\Omega_{\rm k}} \chi(z)\right]/H_0\sqrt{|\Omega_{\rm k}|}, & \Omega_{\rm k} > 0,
\end{array}
\right.
\end{equation}
where the comoving radial distance $\chi$ is
\begin{equation}
\chi(z) = \int_0^z \frac{dz'}{H(z')}.
\end{equation}
The supernovae magnitudes, once standardized, are related
to the distance by
\begin{equation}
m \equiv \mu + M = 5 \log_{10}{d_L}+M + 25,
\end{equation}
where $d_L$ is in units of Mpc. The unknown absolute magnitude $M$ of the supernovae is
a nuisance parameter in the fit and is degenerate with $H_0$.  Hence supernovae
measure relative distances within the set.
In Fig.~\ref{fig:UNION2}, we plot the prediction for the distance modulus from the overall best-fit model with respect to its counterpart from the best-fit $\Lambda$CDM model. The effect of varying parameters in the expansion history is illustrated in Fig.~\ref{fig:UNION}.

\begin{figure}
 \resizebox{\hsize}{!}{\includegraphics{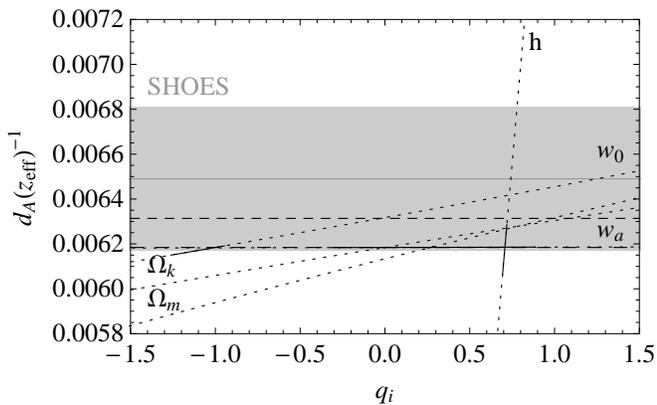}}
 \caption{Inverse angular diameter distance $d_A\left( z_{\rm eff} = 0.04 \right)^{-1}$ for the parameters $q_i \in \{ \Omega_{\rm m}, \Omega_{\rm k}, h, w_0, w_a \}$. The solid segments of the curves correspond to the 1D-marginalized 68\% confidence limits of the respective parameter. Only the according parameter is varied while other parameters are fixed to their best-fit $\Lambda$CDM values. The long-dashed line is the overall best-fit $\Lambda$CDM prediction and the dashed line corresponds to the prediction of the overall best-fit model. Note that due to the low effective redshift $z_{\rm eff}=0.04$, predictions for different values of $w_a$ nearly overlap with the best-fit $\Lambda$CDM prediction. For $\Omega_{\rm m}$ and $\Omega_{\rm k}$ 1D-marginalized 68\% confidence intervals are too narrow to be distinguishable in the plot, i.e., $\Omega_{\rm m}\in(0.255, 0.282)$ and $\Omega_{\rm k}\in(-8.35\times10^{-3}, 2.10\times10^{-3})$, respectively.}
 \label{fig:SHOES}
\end{figure}

The acoustic peaks in the CMB and the measurement of the local Hubble constant additionally provide absolute distance probes, which complement the relative distance measure of the supernovae.
For the Hubble constant, we utilize the SHOES measurement, $H_0=74.2\pm3.6~\rm{km \: s^{-1} \: Mpc^{-1}}$, which employs Cepheid measurements to link the low-redshift supernovae to the distance scale established by the maser galaxy NGC 4258. In the analysis, we use the generalization of this measurement as a constraint on the inverse luminosity distance at $z_{\rm eff} = 0.04$, i.e.,~\cite{reid:09}
\begin{eqnarray}
 d_A\left( z_{\rm eff} \right)^{-1} & = & \left( 1+z_{\rm eff} \right) r\left( z_{\rm eff} \right)^{-1} \nonumber \\
 & = & \left( 1+z_{\rm eff} \right)^2 d_L\left( z_{\rm eff} \right)^{-1} \nonumber \\
 & \simeq & \left( 6.49 \pm 0.32 \right) \times 10^{-3}~\textrm{Mpc}^{-1},
\end{eqnarray}
where $d_A$ is the angular diameter distance and the fiducial cosmology is $h=0.742$, $\Omega_{\rm m} = 0.3$, $\Omega_{\rm k} = 0$, $w_0 = -1$, $w_a = 0$. Figure~\ref{fig:SHOES} demonstrates predictions for $d_A(z_{\rm eff})^{-1}$ and effects on the observable from varying parameters in the expansion history. The gray band corresponds to the $1\sigma$ region of the SHOES measurement, where the solid gray line indicates the mean value.

\begin{figure}
 \resizebox{\hsize}{!}{\includegraphics{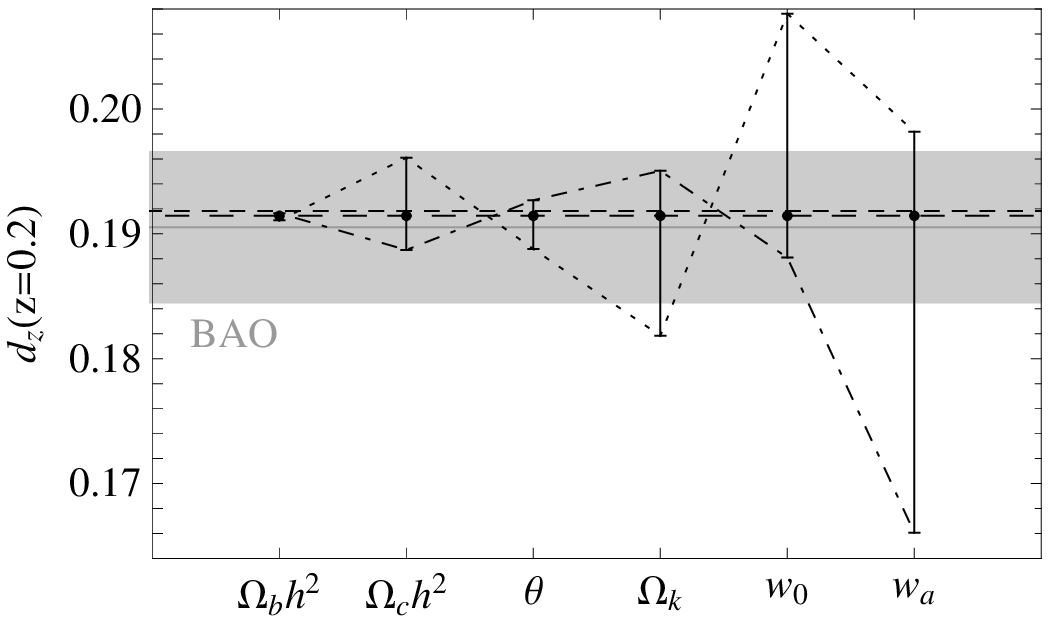}}
 \resizebox{\hsize}{!}{\includegraphics{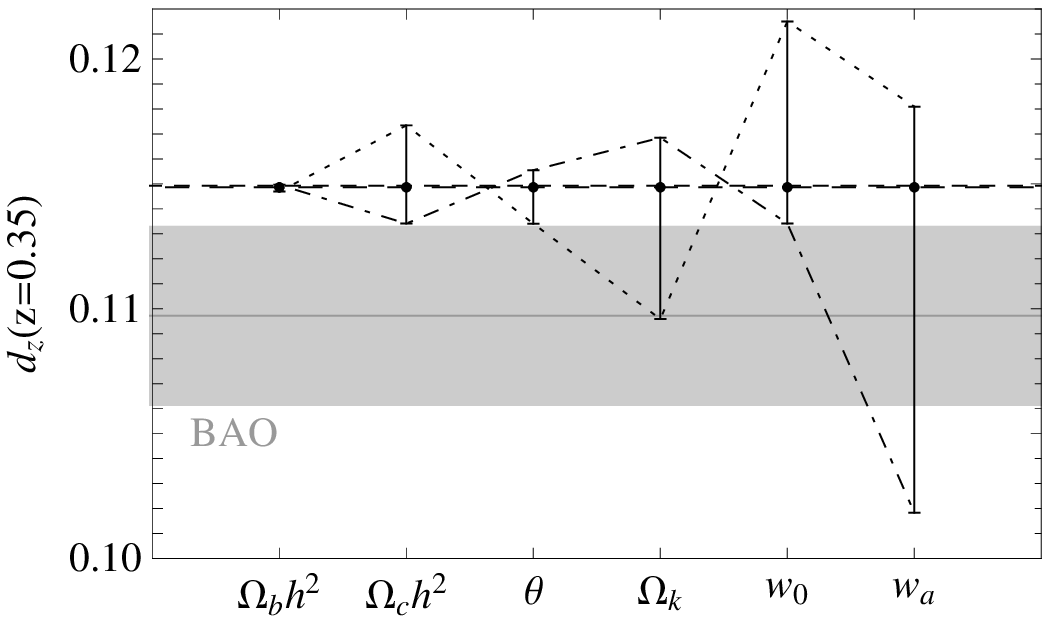}}
 \caption{The ratio $d_z$ of the BAO measurement at redshifts $z=0.2$ (upper panel) and $z=0.35$ (lower panel) for different parameter values. The horizontal dashed and long-dashed line corresponds to the overall best-fit model and best-fit $\Lambda$CDM prediction, respectively. They nearly overlap. The dot-dashed and dotted lines indicate upper and lower 1D-marginalized 68\% confidence limits, respectively. These lines do not have any relevance in between the discrete parameter abscissae and only serve to indicate relative enhancement or suppression due to parameter variations with respect to the best-fit $\Lambda$CDM model.}
 \label{fig:BAO}
\end{figure}

Further, we apply the BAO distance measurement of~\cite{BAO:09} that is obtained from analyzing the clustering of galaxies from the Sloan Digital Sky Survey (SDSS)~\cite{SDSS:09} and the 2-degree Field Galaxy Redshift Survey (2dFGRS)~\cite{2dFGRS:05}, constraining the ratio
\begin{equation}
 d_z \equiv \frac{r_s \left(z_d\right)}{d_V(z)} \equiv \frac{ r_s \left(z_d\right) H(z)^{1/3} }{ (1+z)^{2/3} d_A(z)^{2/3} z^{1/3} }
\end{equation}
at $z=0.2$ and $z=0.35$. Here $r_s \left(z_d\right)$ denotes the comoving sound horizon at the baryon drag epoch $z_d$. In Fig.~\ref{fig:BAO}, we plot predictions for $d_z$ at $z=0.2$ and $z=0.35$ and effects on the observables from varying parameters in the expansion history. The gray bands correspond to the $1\sigma$ region of the BAO distance measurement, where the solid gray lines indicate the mean values.

These probes place tight constraints on the background parameters in Eq.~(\ref{eq:hubbleexpansion}) and help to identify sources for the growth of structure and break degeneracies.

\subsubsection{The cosmic microwave background}

\begin{figure*}
 \resizebox{\hsize}{!}{\includegraphics{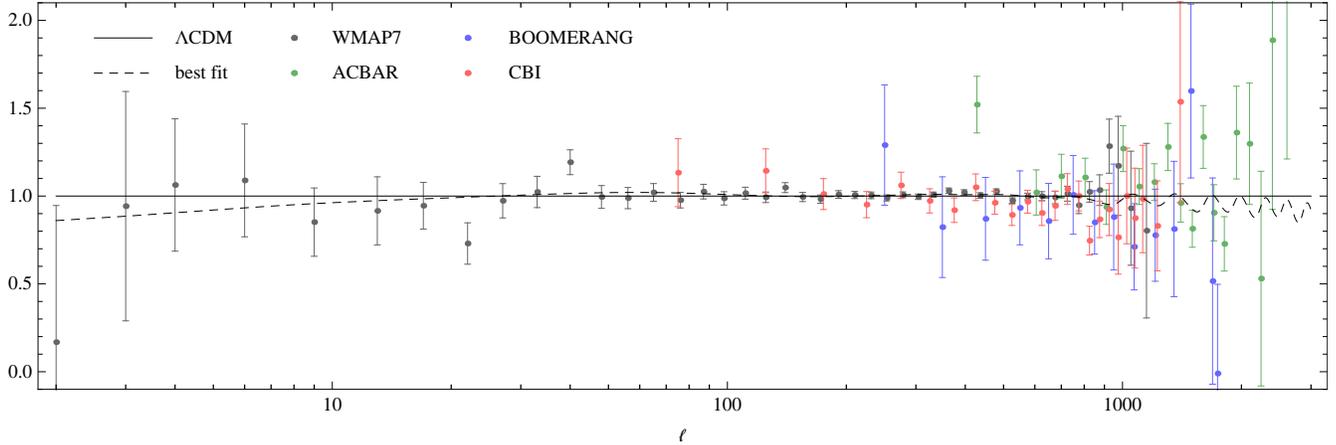}}
 \caption{Ratio of the prediction of the CMB temperature anisotropy power spectrum of the overall best-fit model (dashed line) with respect to its best-fit $\Lambda$CDM counterpart.}
 \label{fig:CMB}
\end{figure*}

\begin{figure*}
 \resizebox{\hsize}{!}{\includegraphics{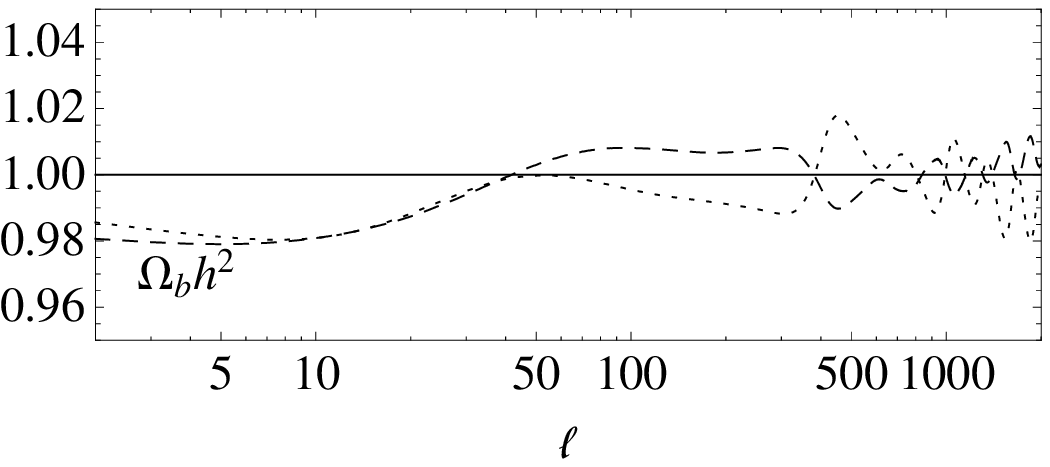}\includegraphics{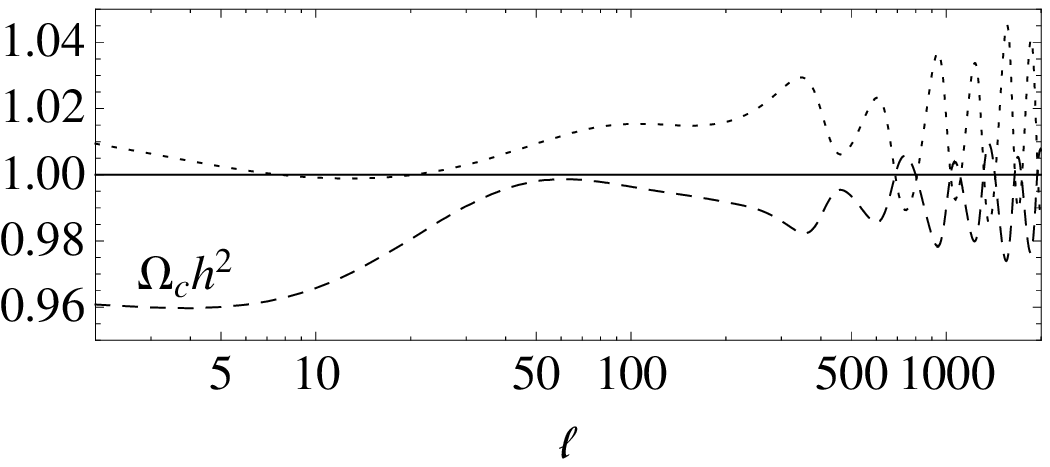}\includegraphics{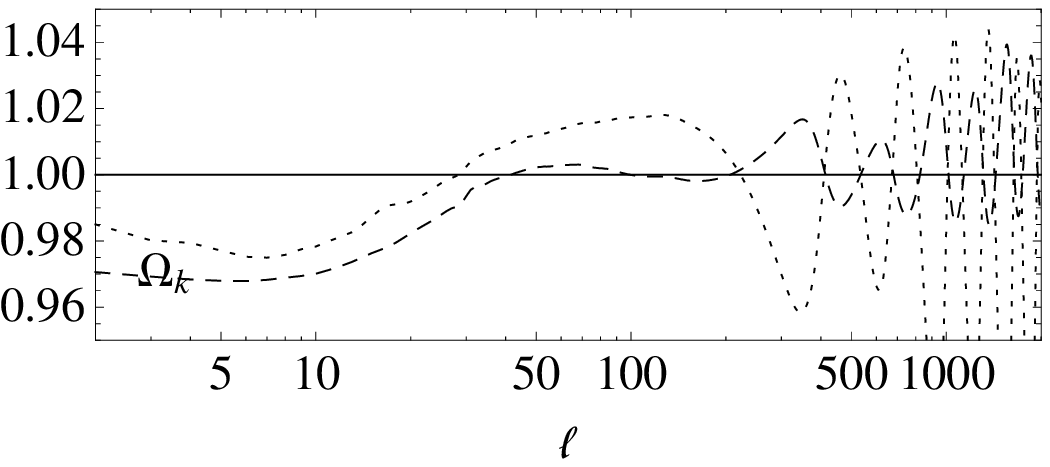}}
 \resizebox{\hsize}{!}{\includegraphics{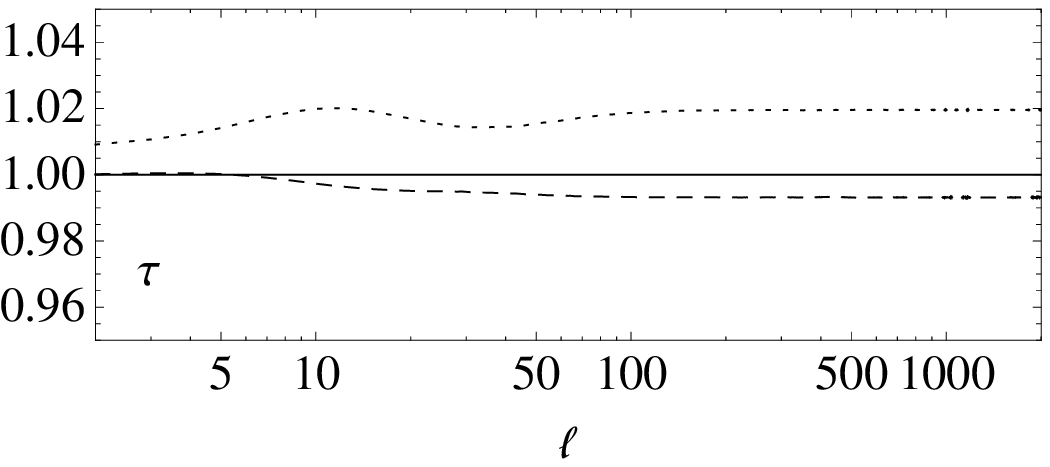}\includegraphics{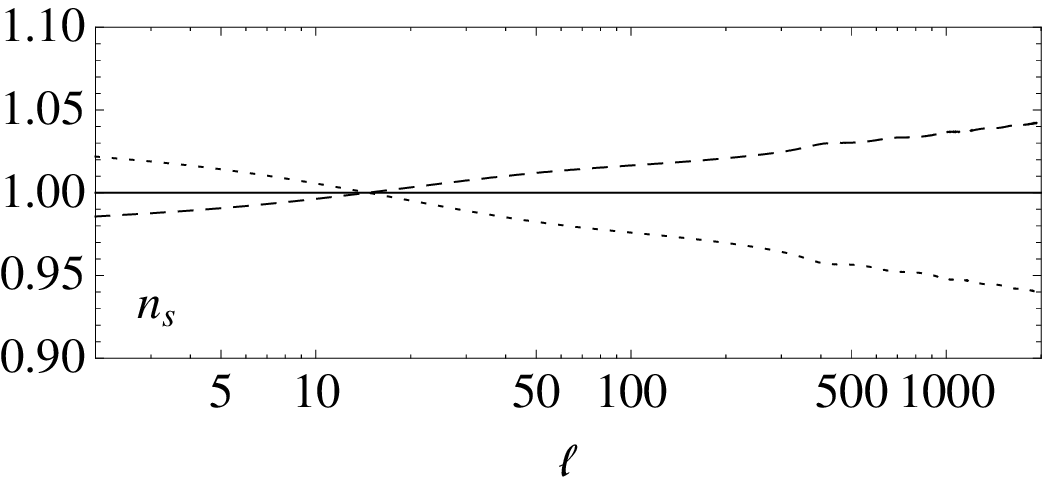}\includegraphics{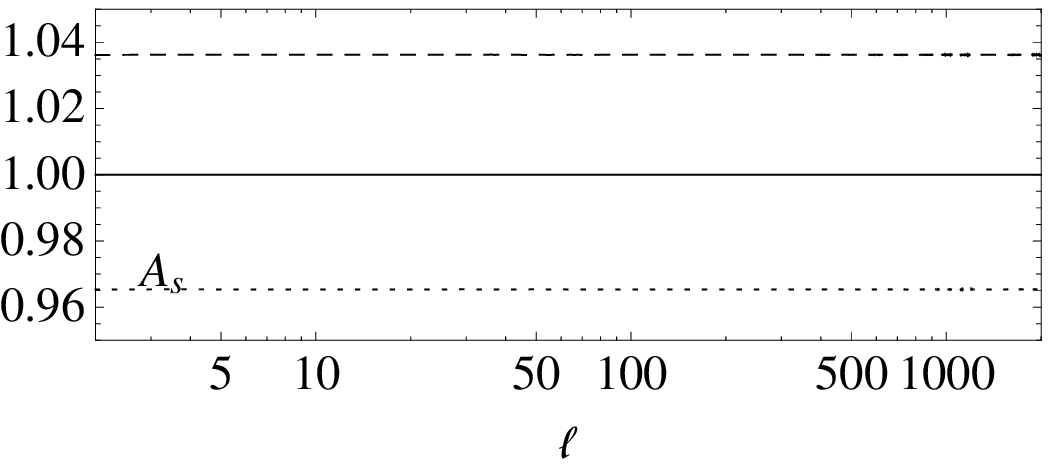}}
 \resizebox{\hsize}{!}{\includegraphics{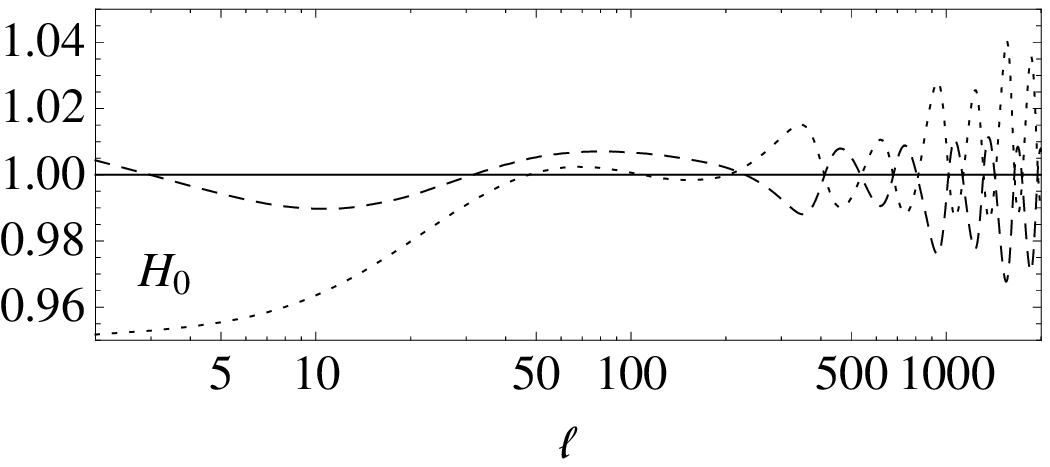}\includegraphics{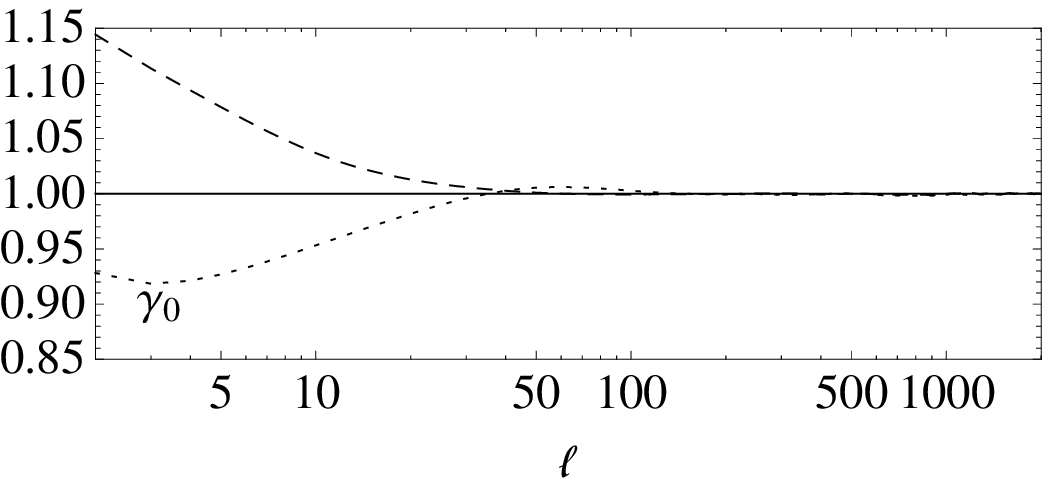}\includegraphics{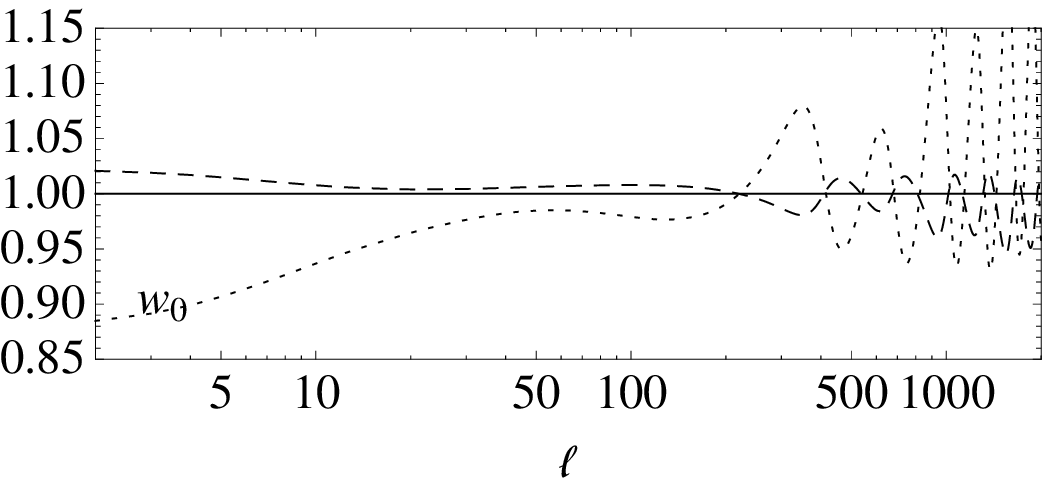}}
 \resizebox{\hsize}{!}{\includegraphics{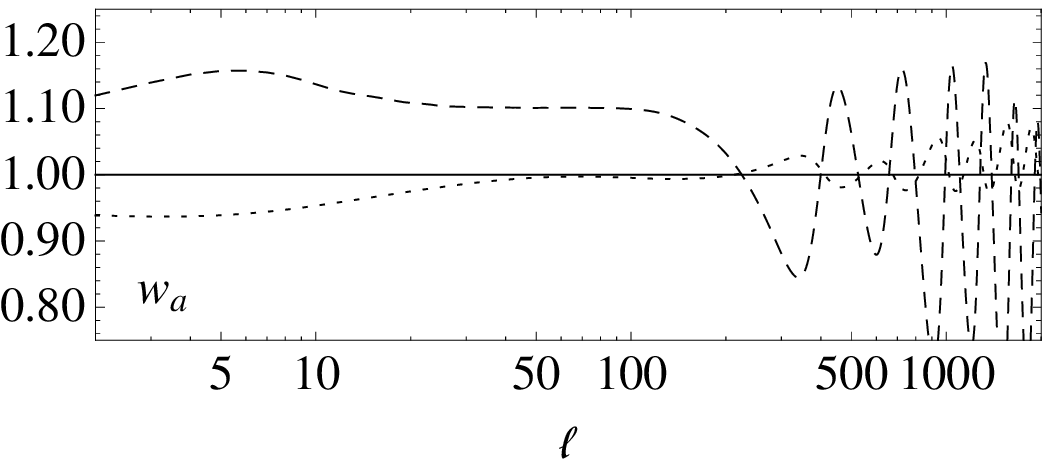}\includegraphics{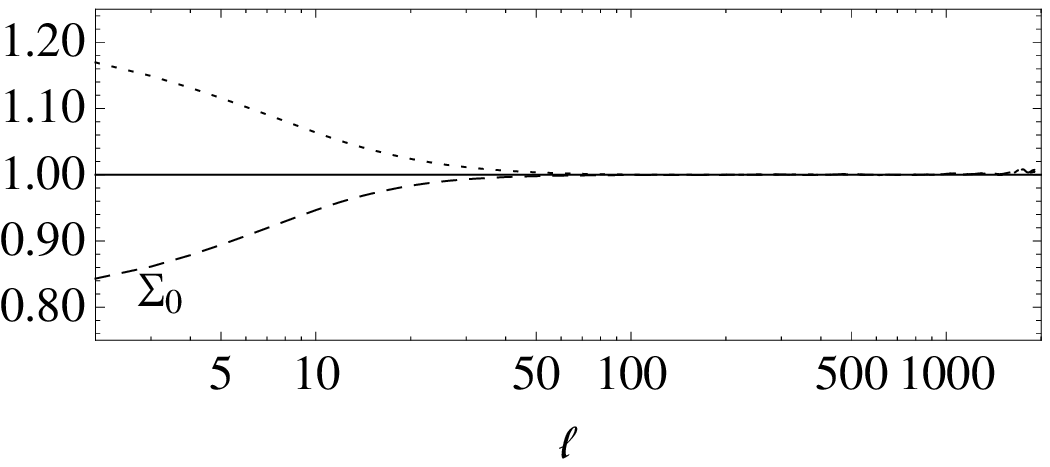}\includegraphics{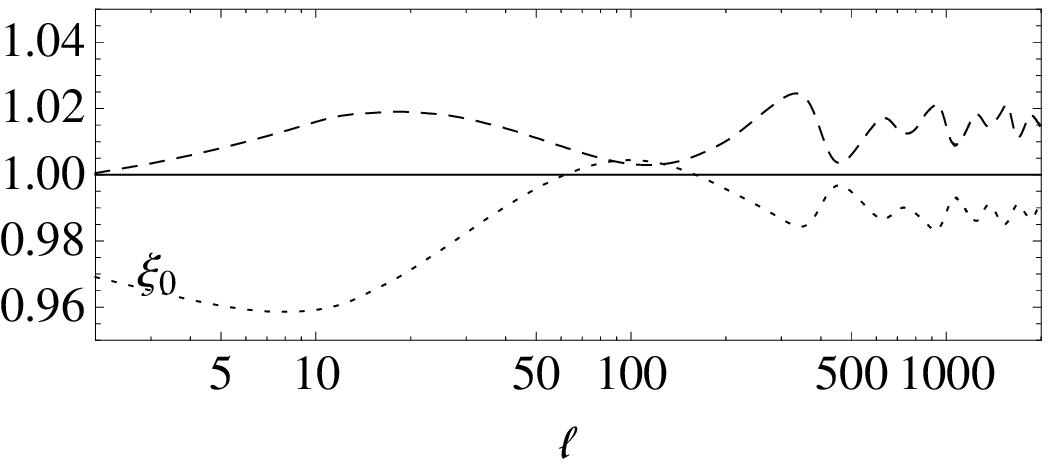}}
 \caption{Ratio of the prediction of the 1D-marginalized 68\% confidence limits of cosmological parameters with respect to the overall best-fit $\Lambda$CDM prediction for the CMB temperature anistropies. Only the respective parameter is varied. Other parameters are fixed to their best-fit $\Lambda$CDM values. Upper limits are indicated by dashed lines, lower limits by dotted lines. Oscillatory behavior at the acoustic peaks indicate shifts of peak position. Note that the first acoustic peak is located at $\ell\simeq220$ and the second at $\ell\simeq540$. Various degeneracies between parameters and sets of parameters are clearly identifiable.}
 \label{fig:RCMB}
\end{figure*}

\begin{figure}
 \resizebox{\hsize}{!}{\includegraphics{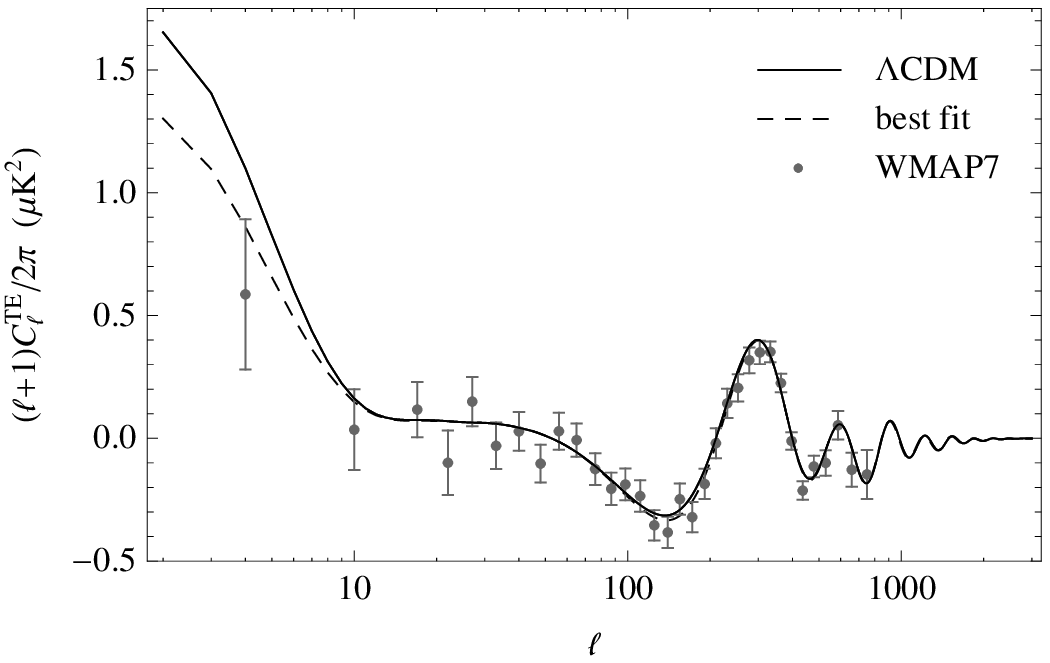}}
 \caption{CMB temperature-polarization cross power spectrum of the overall best-fit model (dashed line) and the best-fit $\Lambda$CDM counterpart (solid line).}
 \label{fig:CMB_TE}
\end{figure}

The CMB probes the geometry of the background history as well as the formation of large-scale structure. The latter manifests itself on the largest scales through the ISW effect from the evolution of the gravitational potential. To predict these effects we connect our parametrization to the PPF formalism and utilize the PPF modifications to CAMB~\cite{CAMB:99} implemented in~\cite{fang:08b}. We configure the PPF parameters as described in Appendix~\ref{sec:ppf_connection}. This connection and the incorporation of the PPF formalism into a standard Einstein-Boltzmann linear theory solver yields an efficient way to obtain predictions from our parametrization for the CMB. It also prevents violations of energy-momentum conservation and avoids gauge artifacts in our results.

In Fig.~\ref{fig:CMB}, we plot the CMB angular temperature anisotropy power spectrum for the overall best-fit model with respect to the prediction of the best-fit $\Lambda$CDM model. Effects from varying chain parameters are illustrated in Fig.~\ref{fig:RCMB}. Figure~\ref{fig:CMB_TE} illustrates the CMB temperature-polarization (TE) cross power spectrum for $\Lambda$CDM and the overall best-fit model. The errors denote diagonal elements of the covariance matrix and include cosmic variance and instrumental noise. The overall best-fit model provides a better fit to all of the CMB data sets. This has to be attributed not only to the relative suppression of the power spectra at large scales with respect to the best-fit $\Lambda$CDM model but also to deviations at large angular multipoles.

\subsubsection{Weak gravitational lensing and galaxy flows}

\begin{figure}
 \resizebox{\hsize}{!}{\includegraphics{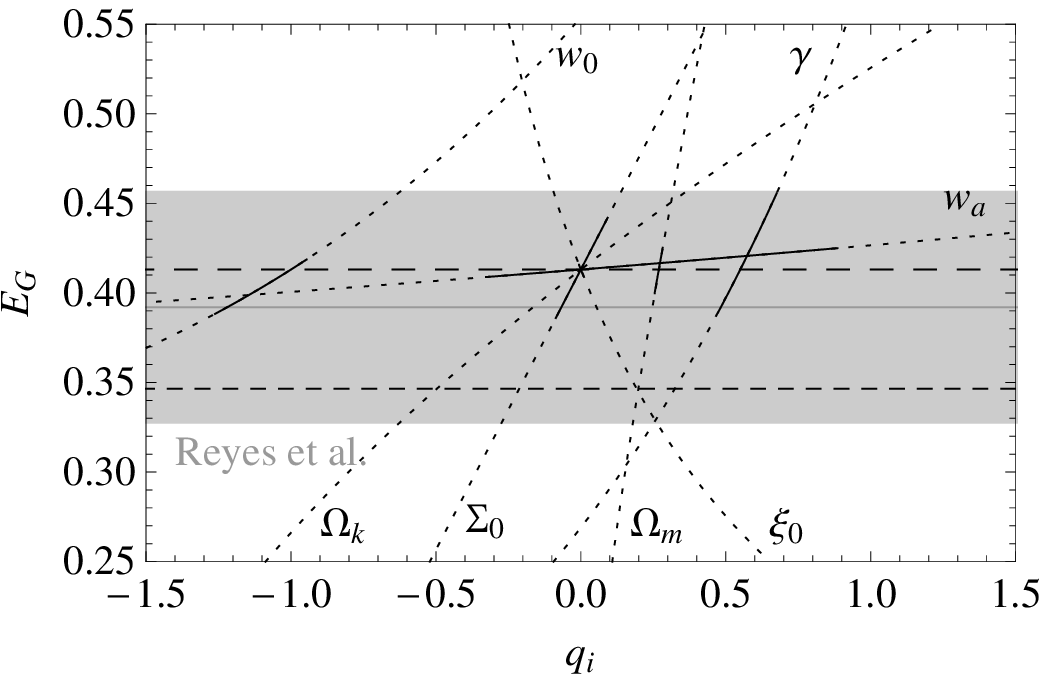}}
 \caption{$E_G$ at $z=0.32$ for the parameters $q_i \in \{ \Omega_{\rm m}, \Omega_{\rm k}, \gamma_0, w_0, w_a, \Sigma_0, \xi_0 \}$. The solid segments of the curves correspond to the 1D-marginalized 68\% confidence limits of the respective parameter. Only the according parameter is varied. Other parameters are fixed to their best-fit $\Lambda$CDM values. The long-dashed line is the overall best-fit $\Lambda$CDM prediction and the dashed line indicates the overall best fit. For $\Omega_{\rm k}$ and $\xi_0$ 1D-marginalized 68\% confidence intervals are too narrow to be distinguishable in the plot, i.e., $\Omega_{\rm k}\in(-8.35\times10^{-3}, 2.10\times10^{-3})$ and $\xi_0\in(-6.41\times10^{-3}, 8.66\times10^{-3})$, respectively.}
 \label{fig:EG}
\end{figure}

The relationship of weak gravitational lensing around galaxies to their large-scale velocities has been proposed as a smoking gun of gravity~\cite{zhang:07}. The advantage of such a probe lies in its insensitivity to galaxy bias and initial matter fluctuations. The expectation value of the ratio of galaxy-galaxy to galaxy-velocity cross correlations of the same galaxies yields an estimator $E_G$~\cite{zhang:07}. We have
\begin{equation}
E_G \equiv \left[ \frac{\nabla^2(\Psi-\Phi)}{3H_0^2a^{-1}\frac{d \ln \Delta_{\rm m}}{d \ln a}\Delta_{\rm m}} \right]_{k=\frac{\ell}{\bar{\chi}},\bar{z}} = \frac{\Omega_{\rm m}}{\Omega_{\rm m}(a)^{\gamma}} \frac{\Sigma}{1+\xi}.
\end{equation}
Recently this quantity has been measured analyzing $70\,205$ luminous red galaxies (LRGs)~\cite{SDSS:01} from the SDSS~\cite{SDSS:00}, yielding $E_G = 0.392\pm0.065$~\cite{reyes:10} at the redshift $z=0.32$ by averaging over scales $R=(10-50)h^{-1}~\textrm{Mpc}$.

Figure~\ref{fig:EG} illustrates predictions for $E_G$ for different parameter values of $q_i \in \{ \Omega_{\rm m}, \Omega_{\rm k}, \gamma_0, w_0, w_a, \Sigma_0, \xi_0 \}$, when varying only one parameter at a time and fixing the others to their overall best-fit $\Lambda$CDM values. 1D-marginalized 68\% confidence limits are obtained from the MCMC analysis in~\textsection\ref{sec:results}.
The gray shaded band corresponds to the $1\sigma$ region of the $E_G$ measurement with the gray solid line being the mean value.

\subsubsection{Galaxy-ISW cross correlations}\label{sec:gISW}

\begin{figure}
 \resizebox{\hsize}{!}{\includegraphics{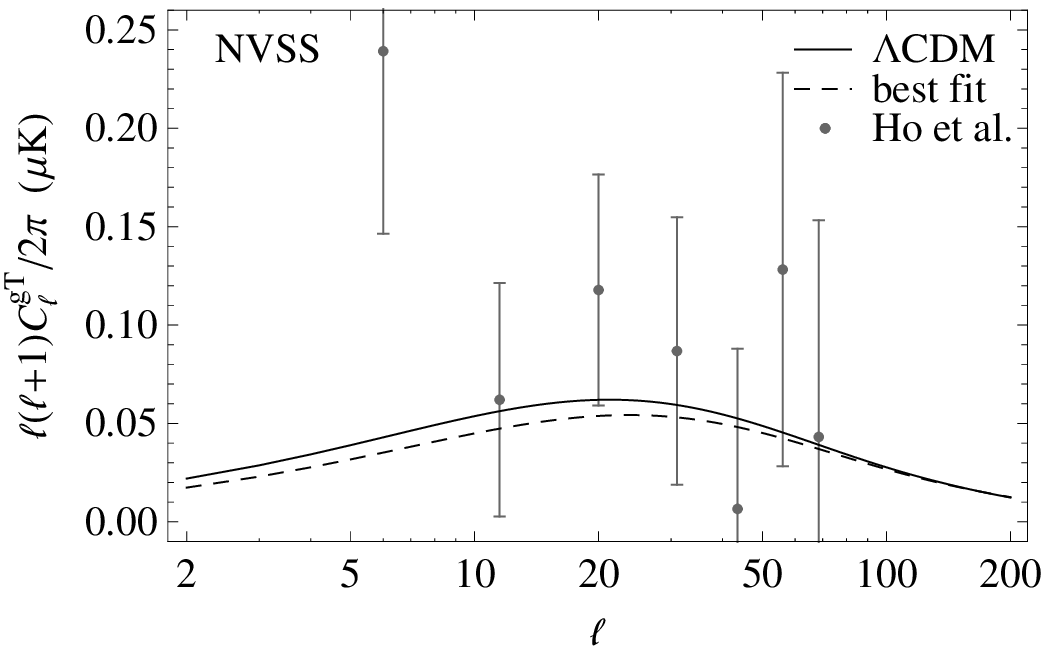}}
 \caption{gISW cross correlations for the NVSS sample. The solid line is the prediction from the best-fit $\Lambda$CDM model, whereas the dashed line corresponds to its counterpart from the overall best-fit model.}
 \label{fig:ISWNVSS}
\end{figure}

\begin{figure*}
 \resizebox{\hsize}{!}{\includegraphics{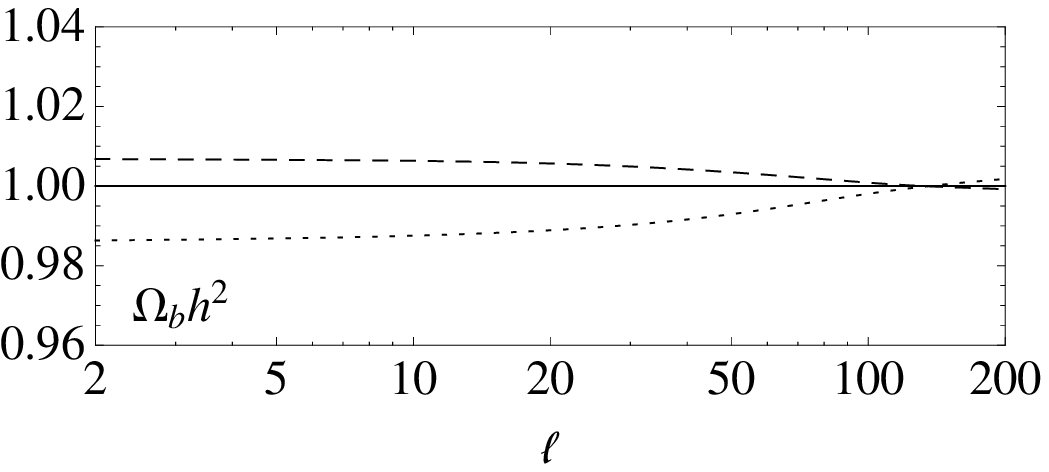}\includegraphics{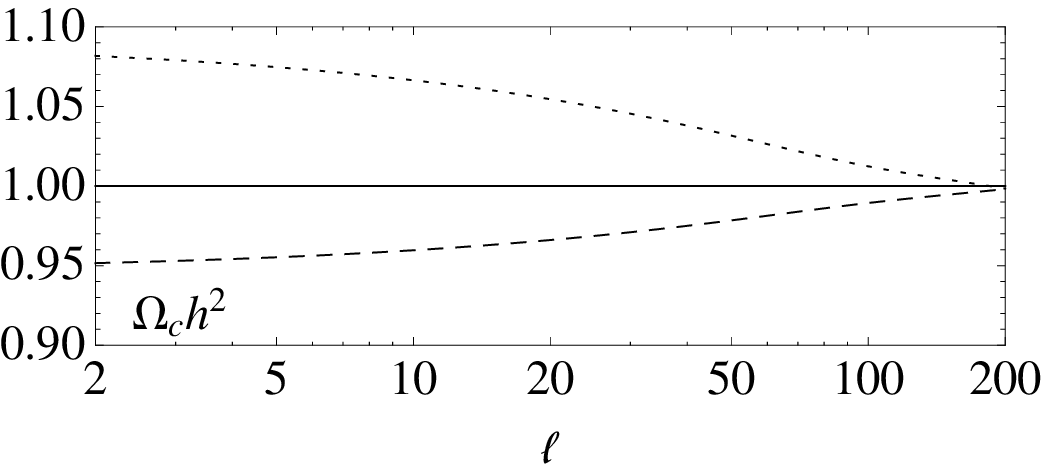}\includegraphics{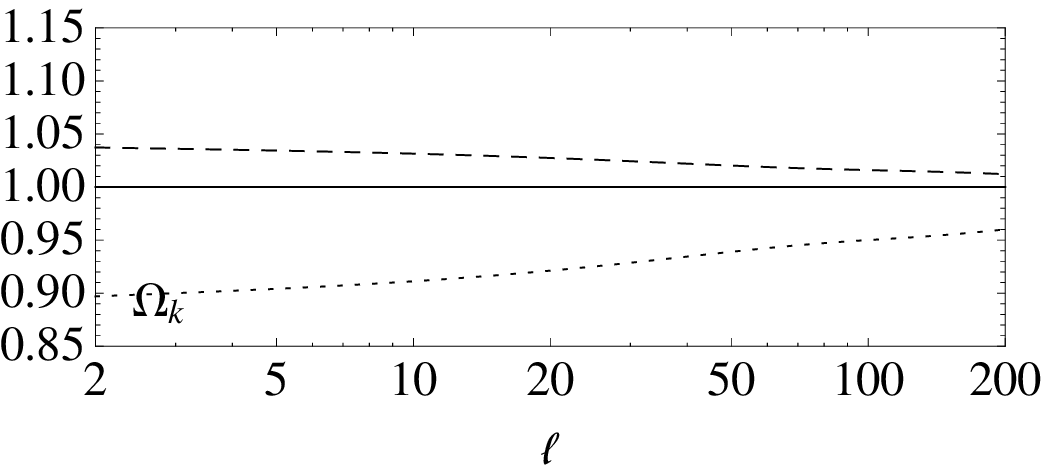}}
 \resizebox{\hsize}{!}{\includegraphics{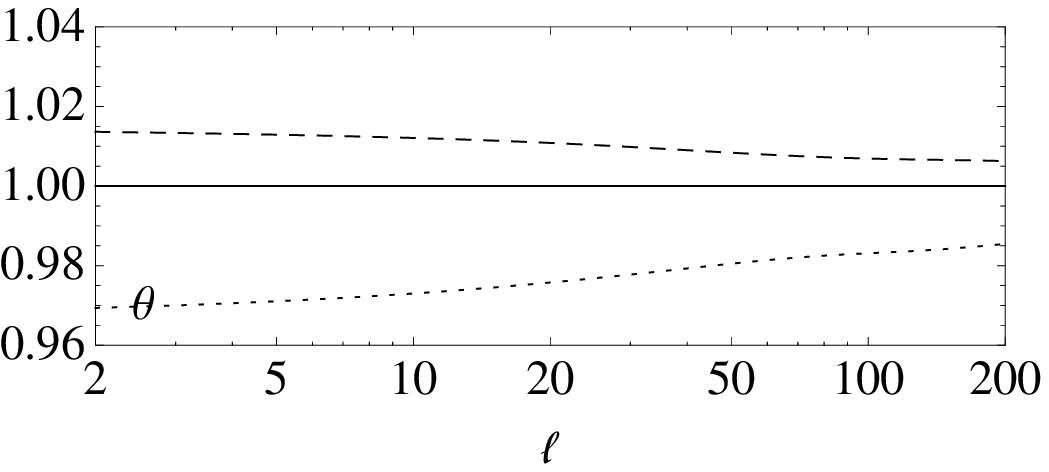}\includegraphics{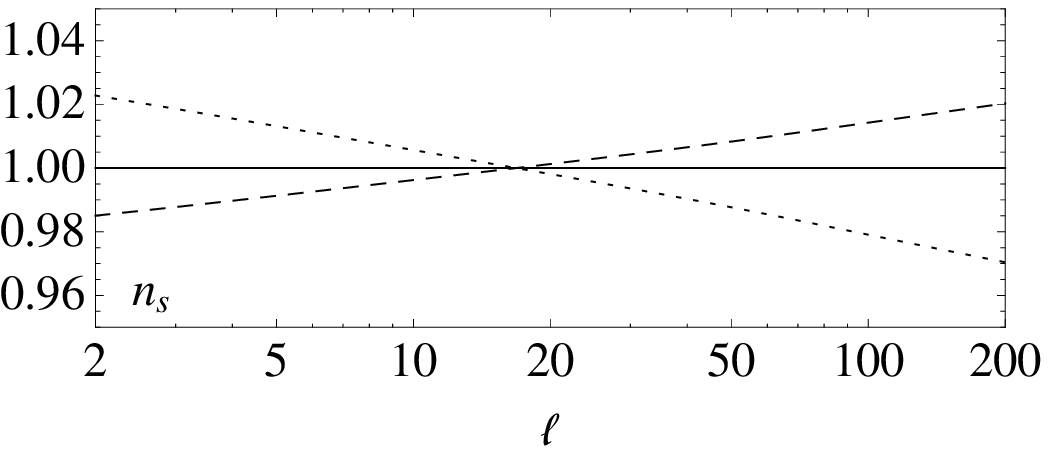}\includegraphics{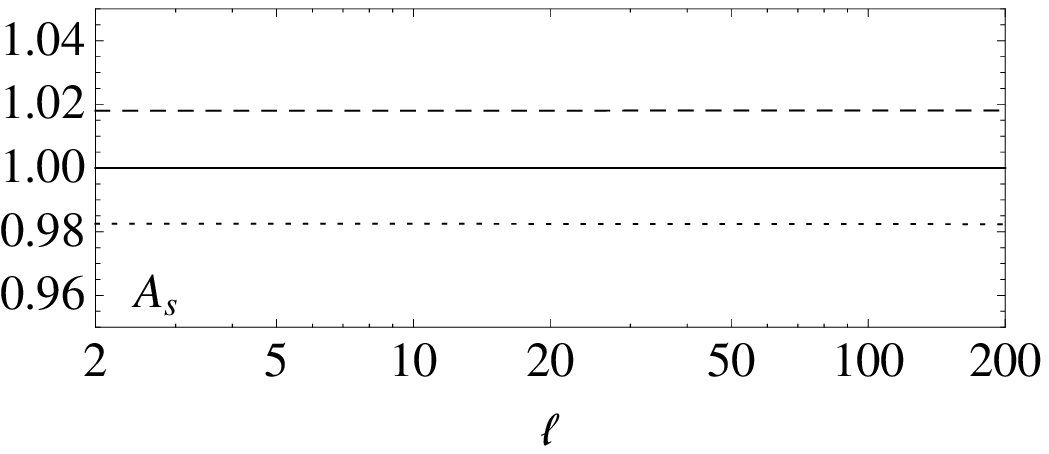}}
 \resizebox{\hsize}{!}{\includegraphics{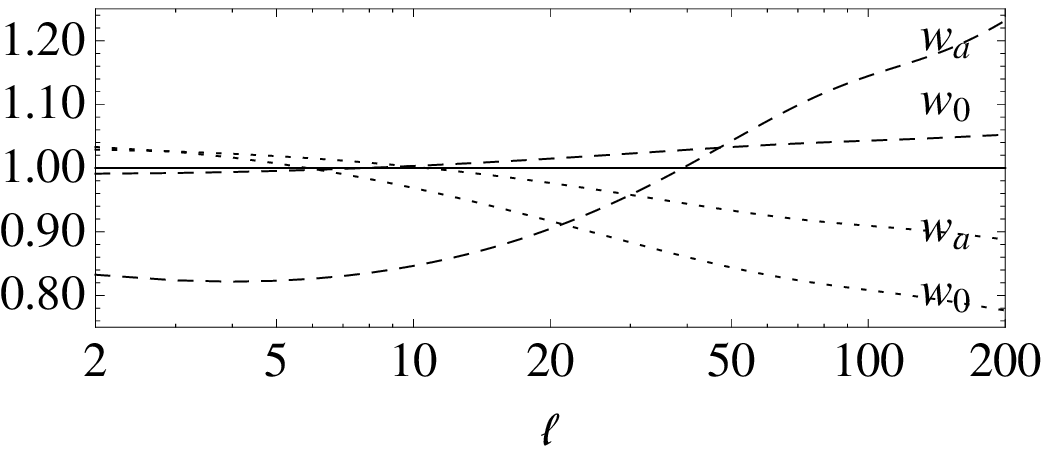}\includegraphics{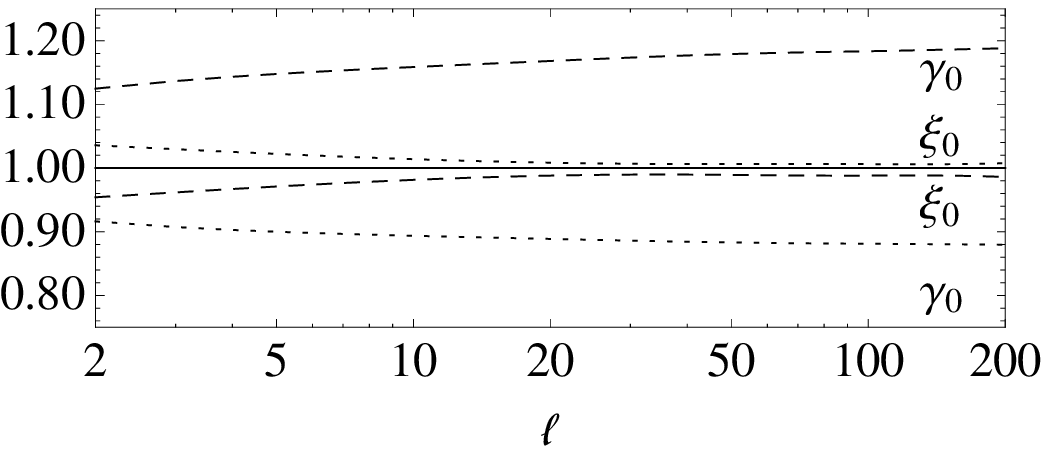}\includegraphics{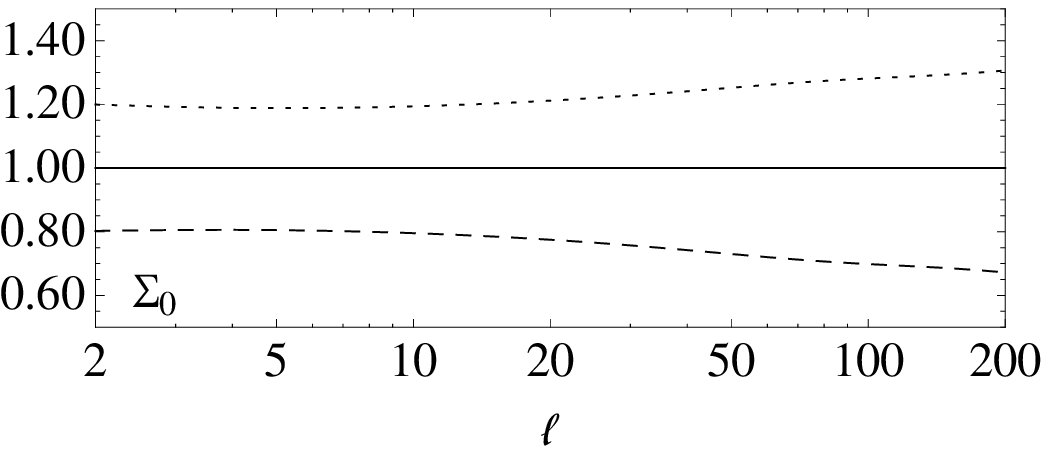}}
 \caption{Ratio of the prediction of the 1D-marginalized 68\% confidence limits of cosmological parameters with respect to the overall best-fit $\Lambda$CDM prediction for the gISW cross correlation in the NVSS sample. Other parameters are fixed to their best-fit $\Lambda$CDM values. Upper limits are indicated by dashed lines, lower limits by dotted lines.}
 \label{fig:RISW}
\end{figure*}

The correlation between galaxy number densities and the CMB anisotropies can be used to isolate the ISW effect in the CMB and has proven to be a useful probe for constraining modifications to standard cosmology (see, e.g.,~\cite{giannantonio:09, lombriser:09, bean:10, lombriser:10}).
We utilize the publicly available {\sc iswwll} code~\cite{ho:08, hirata:08}, where we turn off weak lensing contributions to the likelihood, focusing only on the gISW constraints.
The galaxies in this probe are collected from the Two Micron All Sky Survey (2MASS) extended source catalog (XSC)~\cite{jarrett:00, skrutskie:06}, the LRG samples and photometric quasars (QSO) of the SDSS~\cite{adelman:07}, and the National Radio Astronomy Observatory (NRAO) Very Large Array (VLA) Sky Survey (NVSS)~\cite{condon:98} and correlated with the five-year WMAP~\cite{WMAP:08} CMB anisotropies.
The resulting 42 data points of gISW cross correlations are divided into nine galaxy sample bins $j$ (2MASS0-3, LRG0-1, QSO0-1, and NVSS) based on flux (2MASS) or redshift (LRG and QSO). These data points are a selection of multipole bins from all samples, where the selection is based on the avoidance of nonlinearities and systematic effects from dust extinction, galaxy foregrounds, the thermal Sunyaev-Zel'dovich effect, and point source contamination to affect the gISW cross correlations~\cite{ho:08}.

We evaluate the gISW cross correlation in the Limber and quasistatic approximation, as it is done in the {\sc iswwll} code used for the data analysis. The gISW cross correlation in this approximation reads
\begin{eqnarray}
C_{\ell}^{g_jT} & \simeq & \frac{3 \Omega_{\rm m} H_0^2 T_{\rm CMB}}{(\ell+1/2)^2} \int dz \: f_j(z) H(z) \nonumber\\
& & \times \left[ D \frac{d}{dz}G \right] P \left( \frac{\ell+1/2}{\chi(z)} \right).
\label{eq:ClgT}
\end{eqnarray}
Here, $P(k)$ is the matter power spectrum today. $D$ is the linear density growth rate defined by $\Delta_{\rm m}(k,z) = \Delta_{\rm m}(k,0) D/D|_{z=0}$
and
$G = \Sigma D(1+z)/ \Sigma_i$ is the linear potential growth rate, where $\Sigma_i=\Sigma|_{a=a_i}$ and $a_i\ll1$.

In $\Lambda$CDM, the Limber approximation becomes accurate at the percent level for $\ell\gtrsim10$ and drops approximately as $\ell^2$ at higher $\ell$ (see, e.g.,~\cite{smith:09, loverde:08, afshordi:04}). This condition is satisfied by about $90\%$ of the total 42 data points that are used in the {\sc iswwll} code. The approximation is also valid for, e.g., DGP and $f(R)$ gravity~\cite{lombriser:09, lombriser:10}. Apart from the multipole $\ell$, the error depends also on the width of the redshift distribution, which changes only little with modifications. Given the large errors of the currently available data points at low $\ell$, we conclude that it is safe to apply the Limber approximation and furthermore that it is very useful since it is numerically faster than an exact integration.

The function $f_j(z)$ relates the matter density to the observed projected galaxy overdensity with $f_j(z) = b_j(z)\Pi_j(z)$ in the absence of magnification bias. $\Pi_j(z)$ is the redshift distribution of the galaxies and the bias $b_j(z)$ is assumed independent of scale (cf.~\cite{smith:09}) but dependent on redshift.

The term in brackets in Eq.~(\ref{eq:ClgT}), parametrized by Eqs.~(\ref{eq:Sigma_par}) and (\ref{eq:gamma_par}), can be rewritten as
\begin{equation}
D \frac{d}{dz}G = \left\{ \left[ 1-\Omega_{\rm m}(z)^{\gamma}(1+\xi) \right] \Sigma  + (1+z) \frac{d\Sigma}{dz} \right\} D^2,
\label{eq:growthfactor}
\end{equation}
where
\begin{equation}
D = \exp\left[-\int_0^z \frac{\Omega_{\rm m}(z)^{\gamma}(1+\xi)}{z+1} dz\right].
\end{equation}

For SDSS quasars, the derivation of $f_j(z)$ involves the modified linear growth factor given through Eq.~(\ref{eq:growthfactor}) and magnification bias, for which we use a modification of the lensing window function of~\cite{ho:08}
\begin{equation}
W(z,z') = \frac{3\Sigma H_0^2(1+z)}{2 H(z)} r(z)^2 \left[ \left. \frac{d\ln r(z'')}{d\chi(z'')} \right|_{z'}^{z} \right].
\end{equation}

Note that the gISW analysis uses photometric LRG samples, whereas the $E_G$ measurement is based on spectroscopic LRG samples that do not overlap in redshift. Furthermore, the gISW signal is dominated by large scales and most of the error is caused by sampling variance and shot noise of galaxies. The error on $E_G$ is dominated by uncertainties in lensing and redshift space distortions and most of the signal comes from small scales around $10h^{-1}~\textrm{Mpc}$. Therefore, we can safely neglect correlations between the gISW and $E_G$ data sets.

We plot predictions from the overall best-fit model and the best-fit $\Lambda$CDM model for the gISW cross correlation in the NVSS sample in Fig.~\ref{fig:ISWNVSS} and illustrate effects of varying the chain parameters in Fig.~\ref{fig:RISW}.

\subsection{MCMC likelihood analysis}\label{sec:results}

\begin{table}
\centering
\begin{tabular}{|c|rr|rr|}
\hline
Parameter             & \multicolumn{2}{|c|}{\emph{Flat}} & \multicolumn{2}{|c|}{\emph{Nonflat}} \\
\hline
$\gamma_0$            & $0.57\pm0.10$    & $0.53$         & $0.59\pm0.10$    & $0.60$  \\
$100\:\Omega_{\rm k}$ & \multicolumn{2}{|c|}{$\cdots$}    & $-0.32\pm0.53$   & $-0.81$ \\
\hline
$\gamma_0$            & $0.59\pm0.11$    & $0.56$         & $0.62\pm0.12$    & $0.60$  \\
$w_0$                 & $-1.10\pm0.16$   & $-1.12$        & $-1.05\pm0.18$   & $-1.18$ \\
$w_a$                 & $(-0.19, 1.03)$  & $0.44$         & $(-0.79, 0.99)$  & $0.58$  \\
$100\:\Omega_{\rm k}$ & \multicolumn{2}{|c|}{$\cdots$}    & $-0.54\pm0.77$   & $-0.21$ \\
\hline
$\gamma_0$            & $0.57\pm0.12$    & $0.62$         & $0.58\pm0.11$    & $0.60$  \\
$10\:\Sigma_0$        & $0.02\pm0.93$    & $0.12$         & $-0.03\pm0.95$   & $0.24$  \\
$100\:\Omega_{\rm k}$ & \multicolumn{2}{|c|}{$\cdots$}    & $-0.30\pm0.58$   & $-0.30$ \\
\hline
$\gamma_0$            & $0.58\pm0.10$    & $0.57$         & $0.60\pm0.10$    & $0.56$  \\
$100\:\xi_0$          & $0.11\pm0.76$    & $0.52$         & $0.83\pm1.08$    & $0.47$  \\
$100\:\Omega_{\rm k}$ & \multicolumn{2}{|c|}{$\cdots$}    & $-0.73\pm0.75$   & $-0.59$ \\
\hline
$\gamma_0$            & $0.58\pm0.12$    & $0.55$         & $0.60\pm0.15$    & $0.59$  \\
$w_0$                 & $-1.11\pm0.15$   & $-1.13$        & $-1.05\pm0.18$   & $-1.18$ \\
$w_a$                 & $(-0.99, 1.05)$  & $0.48$         & $(-0.78, 1.13)$  & $0.48$  \\
$10\: \Sigma_0$       & $0.09\pm0.92$    & $-0.08$        & $-0.14\pm0.94$   & $-0.20$ \\
$100\:\Omega_{\rm k}$ & \multicolumn{2}{|c|}{$\cdots$}    & $-0.52\pm0.93$   & $-0.25$ \\
\hline
$\gamma_0$            & $0.57\pm0.12$    & $0.51$         & $0.60\pm0.11$    & $0.56$  \\
$10\:\Sigma_0$        & $0.07\pm0.97$    & $-0.27$        & $0.01\pm0.94$    & $0.18$  \\
$100\:\xi_0$          & $0.15\pm0.80$    & $0.33$         & $0.79\pm1.07$    & $0.57$  \\
$100\:\Omega_{\rm k}$ & \multicolumn{2}{|c|}{$\cdots$}    & $-0.71\pm0.77$   & $-0.47$ \\
\hline
$\gamma_0$            & $0.60\pm0.11$    & $0.67$         & $0.63\pm0.12$    & $0.60$  \\
$w_0$                 & $(-1.31, -1.03)$ & $-1.13$        & $(-1.30, -0.93)$ & $-1.29$ \\
$w_a$                 & $(0.36, 1.45)$   & $1.17$         & $(-0.46, 1.36)$  & $1.29$  \\
$100\:\xi_0$          & $(-1.46, 3.17)$  & $10.06$        & $(-0.58, 2.16)$  & $3.72$  \\
$100\:\Omega_{\rm k}$ & \multicolumn{2}{|c|}{$\cdots$}    & $-0.64\pm0.93$   & $0.23$  \\
\hline
$\gamma_0$            & $0.60\pm0.11$    & $0.60$         & $0.61\pm0.13$    & $0.66$  \\
$w_0$                 & $(-1.29, -1.01)$ & $-1.15$        & $(-1.30, -0.94)$ & $-1.24$ \\
$w_a$                 & $(0.37, 1.43)$   & $1.17$         & $(-0.37, 1.42)$  & $1.26$  \\
$10\:\Sigma_0$        & $0.27\pm0.99$    & $-0.38$        & $0.06\pm0.96$    & $0.50$  \\
$100\:\xi_0$          & $(-1.29, 3.14)$  & $6.33$         & $(-0.67, 2.35)$  & $7.19$  \\
$100\:\Omega_{\rm k}$ & \multicolumn{2}{|c|}{$\cdots$}    & $-0.54\pm1.02$   & $-0.41$ \\
\hline
\end{tabular}
\caption{Mean, standard deviations, and best-fit values for the extra degrees of freedom in the $\gamma [w] [\Sigma] [\xi] [{\rm k}]$ models from using the WMAP7, ACBAR, B03, CBI, UNION2, BAO, SHOES, gISW, and $E_G$ data sets. Values in brackets denote the 1D-marginalized 68\% minimal credible intervals~\cite{hamann:07} quoted for parameter directions in the posterior distribution with distinctive skewness.}
\label{tab:results_one}
\end{table}

\begin{table*}
\centering
\begin{tabular}{|c|rr|rr|rr|}
\hline
Parameter                        & \multicolumn{2}{|c|}{$\Lambda$CDM} & \multicolumn{2}{|c|}{$\gamma w \xi$} & \multicolumn{2}{|c|}{(A)}    \\
\hline
$100\: \Omega_b h^2$             & $2.229\pm0.050$   & $2.244$        & $2.237\pm0.056$   & $2.246$          & $2.231\pm0.056$   & $2.196$  \\
$\Omega_c h^2$                   & $0.1114\pm0.0030$ & $0.1121$       & $0.1066\pm0.0075$ & $0.0995$         & $0.1058\pm0.0077$ & $0.0935$ \\
$\theta$                         & $1.0397\pm0.0024$ & $1.0406$       & $1.0443\pm0.0063$ & $1.0510$         & $1.0453\pm0.0068$  & $1.0560$ \\
$\tau$                           & $0.085\pm0.014$   & $0.087$        & $0.086\pm0.015$   & $0.090$          & $0.086\pm0.015$   & $0.091$  \\
$n_s$                            & $0.960\pm0.012$   & $0.962$        & $(0.947, 0.991)$  & $1.016$          & $(0.948, 0.992)$  & $1.011$  \\
$\ln \left[ 10^{10} A_s \right]$ & $3.199\pm0.036$   & $3.199$        & $(3.034, 3.270)$  & $2.810$           & $(3.009, 3.271)$  & $2.744$  \\
$\gamma_0$                       & \multicolumn{2}{|c|}{\ldots}       & $0.60\pm0.11$     & $0.67$           & \multicolumn{2}{|c|}{\ldots} \\
$w_0$                            & \multicolumn{2}{|c|}{\ldots}       & $(-1.31, -1.03)$  & $-1.13$          & $(-1.29, -1.02)$  & $-1.07$  \\
$w_a$                            & \multicolumn{2}{|c|}{\ldots}       & $(0.36, 1.45)$    & $1.17$           & $(0.38, 1.47)$    & $1.09$   \\
$100\: \xi_0$                    & \multicolumn{2}{|c|}{\ldots}       & $(-1.46, 3.17)$   & $10.06$          & $(-1.42, 3.63)$   & $9.67$   \\
\hline
$\Omega_{\rm m}$                 & $0.268\pm0.014$   & $0.269$        & $0.258\pm0.018$   & $0.233$          & $0.256\pm0.019$   & $0.224$  \\
$\sigma_8$                       & $0.806\pm0.021$   & $0.812$        & $0.779\pm0.077$   & $0.749$          & $0.800\pm0.063$   & $0.769$  \\
$H_0$                            & $70.6\pm1.3$      & $70.7$         & $70.7\pm1.7$      & $72.3$           & $70.7\pm1.8$      & $71.8$   \\
\hline
\end{tabular}
\caption{The same as in Table~\ref{tab:results_one}, but for the $\Lambda$CDM, $\gamma w \xi$, and (A) models and including the basic cosmological parameters along with matter density, power spectrum normalization $\sigma_8$, and the Hubble constant.}
\label{tab:results_two}
\end{table*}

\begin{table*}
\centering
\begin{tabular}{|c|rr|rr|rr|}
\hline
Parameter                        & \multicolumn{2}{|c|}{(B)}    & \multicolumn{2}{|c|}{(C)}    & \multicolumn{2}{|c|}{(C)$^*$} \\
\hline
$100\: \Omega_b h^2$             & $2.236\pm0.056$   & $2.230$  & $2.229\pm0.056$   & $2.189$  & $2.232\pm0.055$   & $2.201$  \\
$\Omega_c h^2$                   & $0.1097\pm0.0059$ & $0.1150$ & $0.1041\pm0.0073$ & $0.0976$ & $0.1023\pm0.0067$ & $0.0986$ \\
$\theta$                         & $1.0413\pm0.0048$ & $1.0395$ & $(1.0474, 1.0559)$& $1.0550$ & $1.0489\pm0.0066$ & $1.0544$ \\
$\tau$                           & $0.085\pm0.014$   & $0.086$  & $0.086\pm0.015$   & $0.081$  & $0.087\pm0.014$   & $0.085$  \\
$n_s$                            & $0.963\pm0.014$   & $0.959$  & $(0.953, 1.002)$  & $0.996$  & $(0.970, 1.018)$  & $0.998$  \\
$\ln \left[ 10^{10} A_s \right]$ & $3.179\pm0.059$   & $3.221$  & $(2.830, 2.958)$  & $2.841$  & $(2.775, 2.982)$  & $2.864$  \\
$w_a$                            & $(-0.23, 0.53)$   & $-0.20$  & \multicolumn{2}{|c|}{\ldots} & \multicolumn{2}{|c|}{\ldots} \\
$100\: \xi_0$                    & $(-0.81, 1.41)$   & $-0.18$  & $(5.07, 7.64)$    & $7.88$   & $(4.19, 8.59)$    & $7.68$   \\
$\lambda_0$                      & \multicolumn{2}{|c|}{\ldots} & $(0.31, 1.22)$    & $1.02$   & $(0.71, 1.25)$    & $1.02$   \\
\hline
$\Omega_{\rm m}$                 & $0.267\pm0.015$   & $0.274$  & $0.253\pm0.021$   & $0.227$  & $0.248\pm0.021$   & $0.232$  \\
$\sigma_8$                       & $0.809\pm0.060$   & $0.829$  & $0.811\pm0.062$   & $0.818$  & $0.788\pm0.062$   & $0.805$  \\
$H_0$                            & $70.4\pm1.7$      & $70.9$   & $70.7\pm1.8$      & $72.5$   & $71.0\pm2.0$      & $72.0$   \\
\hline
\end{tabular}
\caption{The same as in Table~\ref{tab:results_one}, but for the (B) and (C) models, where (C)$^*$ denotes the case where no gISW data are used. The best-fit values quoted here for the (C) model define the overall highest maximum-likelihood model found for the combination of all of the data sets studied in this paper and are used to derive predictions for illustrations in~\textsection{\ref{sec:predictions}}. For the (C) models, the 1D-marginalized minimal credible intervals correspond to the mode with larger mean likelihood.}
\label{tab:results_three}
\end{table*}

\begin{table*}
\centering
\begin{tabular}{|c|crrrrrrl|r|}
\hline
Statistic & $\Lambda$CDM & $\gamma w \xi$ & $\gamma w \xi {\rm k}$ & $\gamma w \Sigma \xi$ & $\gamma w \Sigma \xi {\rm k}$ & (A)    & (C)    & \emph{Others} & (C)$^*$ \\
\hline
$-2 \Delta \ln \mathcal{L}_{\rm max}$
          & $0$          & $-5.2$         & $-2.2$                 & $-3.8$                & $-3.4$                        & $-4.9$ & $-7.3$ & $>-0.5$       & $-6.8$ \\
$-2 \Delta \langle \ln \mathcal{L} \rangle_{\rm s}$
          & $0$          & $0.9$          & $2.9$                  & $2.2$                 & $3.4$                         & $0.0$  & $-0.9$  & $>0$         & $-2.9$ \\
$\overline{\langle \mathcal{L}\rangle_{\rm s}}$
          & $1$          & $2.6$          & $0.6$                  & $1.0$                 & $0.6$                         & $2.3$  & $6.5$  & $<1$          & $11.3$ \\
\hline
\end{tabular}
\caption{Comparison of the goodness of fit of the models with respect to $\Lambda$CDM, where $\ln \mathcal{L}_{\rm max}^{\Lambda\textrm{CDM}}=-4049.6$, $\langle \ln \mathcal{L}_{\Lambda\textrm{CDM}} \rangle_{\rm s}=-4052.8$, and $\ln \langle \mathcal{L}_{\Lambda\textrm{CDM}} \rangle_{\rm s}=-4051.9$. $\overline{\langle \mathcal{L}\rangle_{\rm s}}>1$ suggests that the adoption of the extra parameters might be justified. (C)$^*$ denotes the case where no gISW data are used.}
\label{tab:stats}
\end{table*}

We turn to the likelihood analysis of the cosmological parameter space. As our elementary set of parameters we use $\mathcal{C} = \left\{ \Omega_bh^2, \Omega_ch^2, \theta, \tau, n_s, \ln \left[ 10^{10} A_s \right], \gamma_0 \right\}$, where for the concordance model $\mathcal{C}_{\Lambda\textrm{CDM}} \approx \mathcal{C} \cap \{ \gamma_0 = 0.55 \}$. We implement the following flat priors: $\Omega_bh^2 \in (0.01,0.1)$, $\Omega_ch^2 \in (0.045,0.99)$, $\theta \in (0.5,10)$, $\tau \in (0.01,0.8)$, $n_s \in (0.5,1.5)$, $\ln \left[ 10^{10} A_s \right] \in (2.7,4)$~\footnote{Parameter constraints on $\ln \left[ 10^{10} A_s \right]$ are close to its prior boundary, suggesting a non-negligible influence of the prior. We have checked that using a flat top-hat prior with the lower boundary set at 2 instead of 2.7 does not alter our results in a significant manner.}, and $\gamma_0 \in (-5,5)$. In addition we use all the sets of combinations between the phenomenological parameters and spatial curvature $p_i \in \mathcal{P}$, where $\mathcal{P} = \{W, \Sigma_0, \xi_0, \Omega_{\rm k} \}$ and $W = \{w_0, w_a\}$, being free and taking on their $\Lambda$CDM values $\{w_0=-1, w_a=0, \Sigma_0=0, \xi_0=0, \Omega_{\rm k}=0 \}$. We assign the following flat priors to them: $w_0 \in (-5,5)$, $w_a \in (-10,10)$, $\Sigma_0 \in (-5,5)$, $\xi_0 \in (-0.5,0.5)$, and $\Omega_{\rm k} \in (-0.3,0.3)$. As starting centers for $p_i \in \mathcal{P}$ we use the $\Lambda$CDM values. For the (C) model, we use $\lambda_0 \in (-20,20)$ with starting center $\lambda_0=0.5$.

The {\sc cosmomc} package employs the Metropolis-Hastings algorithm~\cite{metropolis:53, hastings:70} for the sampling and the Gelman and Rubin statistic $R$~\cite{gelman:92} for testing the convergence. We require $R-1 < 0.01$ for our runs with two or fewer extra degrees of freedom and $R-1 < 0.02$ for three or more extra degrees of freedom~\footnote{Note that the $\gamma w \Sigma \xi [k]$ models did not acquire the desired accuracy and reached only a convergence of $R-1 \lesssim 0.1$.}.

The chain samples are used to infer marginalized probabilities and mean likelihoods of the posterior. The marginalized distribution is obtained from projecting the samples to the reduced dimensions of a subspace, ignoring information about the goodness of fit and skewness of the distribution in the marginalized directions. Averaging the likelihood for each point of a subspace produces the mean likelihood. If the two curves do not overlap, the distribution is not Gaussian or the priors are not flat. It is important to note that in this case, marginalized probabilities may be amplified due to a larger parameter volume rather than by a better fit to the data. Furthermore, for skew distributions, we quote our 1D-marginalized constraints in terms of minimal credible intervals~\cite{hamann:07} rather than by marginalized confidence limits. The former indicate where the tails occupy equal fractions of the probability distribution, whereas the latter are constructed in such a way that any point inside of the interval has a larger posterior than a point outside of it, being the more meaningful selection in the case of skew or multimodal distributions.

We summarize our results in Table~\ref{tab:results_one} through \ref{tab:stats}.
For comparison of the goodness of fit between the different models, we quote
\begin{eqnarray}
-2 \Delta \ln \mathcal{L}_{\rm max} & = & 2 \ln \left( \mathcal{L}_{\rm max}^{\rm \Lambda CDM}/\mathcal{L}_{\rm max} \right), \label{eq:goodness_a} \\
-2 \Delta \langle \ln \mathcal{L} \rangle_{\rm s} & = & 2\left( \langle \ln \mathcal{L}_{\rm \Lambda CDM}\rangle_{\rm s} - \langle \ln \mathcal{L}\rangle_{\rm s} \right), \\
\overline{\langle \mathcal{L}\rangle_{\rm s}} & = & \langle \mathcal{L}\rangle_{\rm s}/\langle \mathcal{L}_{\rm \Lambda CDM}\rangle_{\rm s} \label{eq:goodness_b}
\end{eqnarray}
in Table~\ref{tab:stats}, where $\mathcal{L}_{\rm max}$ is the maximum likelihood of a model and $\langle \cdot \rangle_{\rm s}$ denotes the average over the Monte-Carlo samples.
Note that MCMC sampling may not provide very accurate best-fit estimates (see, e.g.,~\cite{COSMOMC:02}). The best-fit values may also have a much higher likelihood than the mean, but simultaneously be confined to a very small region of the parameter space. We also give the mean likelihood of the samples $\overline{\langle \mathcal{L}\rangle_{\rm s}}$, which corresponds to taking the posterior distribution as prior in calculating the evidence and in contrary to the maximum likelihood is a quantity that penalizes fine-tuning.
If $\overline{\langle \mathcal{L}\rangle_{\rm s}}$ is greater than unity, this usually suggests that, on average, the extra parameters improve the goodness of fit to the data. This, however, has to be interpreted only as a rule of thumb (see, e.g.,~\cite{COSMOMC:02}).
We only give constraints on the parameters $c_i \in C \backslash \{\gamma_0\}$ for the $\gamma w \xi$, (A), (B), and (C) models in addition to $\Lambda$CDM (see Tables~\ref{tab:results_one} through \ref{tab:results_three}). These models, except for the (B) model, exhibit an improved goodness of fit in terms of maximum and averaged likelihoods over the $\Lambda$CDM model.
We numerically evaluate Eqs.~(\ref{eq:goodness_a}) through (\ref{eq:goodness_b}) and quote the numbers for the $\gamma w [\Sigma] \xi [k]$,  (A), and (C) models in Table~\ref{tab:stats}. For all other models, $\overline{\langle \mathcal{L}\rangle_{\rm s}}$ is smaller than unity, $-2 \Delta \langle \ln \mathcal{L} \rangle_{\rm s}$ is positive, and $-2 \Delta \ln \mathcal{L}_{\rm max}>-0.5$.

Under the assumptions made in \textsection\ref{sec:parametrization}, our results show that constraints on the growth index parameter $\gamma_0$ are only weakly affected by the introduction of the other extra degrees of freedom with consistency of $\Lambda$CDM, i.e., $\gamma_0\approx0.55$, at the 68\% confidence level. The constraints on $\gamma_0$ are dominated by the gISW data and are competitive to existing results in the literature derived from linear and nonlinear probes (cf., e.g.,~\cite{rapetti:08, rapetti:09, bean:10, dossett:10}).

In every scenario studied here, the standard values of all extra parameters, corresponding to the concordance model, lie within their 1D-marginalized 95\% confidence limits. Thus, we conclude that the $\Lambda$CDM model is consistent with the joint set of WMAP7, ACBAR, B03, CBI, UNION2, BAO, SHOES, gISW, and $E_G$ data under the assumption of the existence of extra degrees of freedom of the kind described in~\textsection\ref{sec:parametrization} and form of priors given above. 
Note that constraints on a parameter direction depend a lot on the prior assumed for it or for the parameters used to derive it (see, e.g.,~\cite{COSMOMC:02}). Identifying adequate priors on extra degrees of freedom in the gravitational dynamics based on theoretical contemplations is the object of current research (see, e.g.,~\cite{ferreira:10, song:10}).

For the maximum-likelihood $\Lambda$CDM model we obtain a fit of $-2 \ln \mathcal{L}_{\rm max}^{\rm \Lambda CDM}=8099.3$ using all of the data described above. The overall best-fit to the data was obtained in the chains of model (C) with $-2 \Delta \ln \mathcal{L}_{\rm max} = -7.3$ with respect to the best-fit $\Lambda$CDM model. Note that in principle the best-fit values of model (C) are also contained in the parameter spaces of the $\gamma w [\Sigma] \xi [{\rm k}]$ and (A) models. The fact that none of these models accessed the best-fit point can be attributed to sampling errors in the chains. However, for the maximum-likelihood values of the $\gamma w [\Sigma] \xi [{\rm k}]$ and (A) models, we also obtain increases in fit of $-2 \Delta \ln \mathcal{L}_{\rm max}\lesssim-2$ over the best-fit $\Lambda$CDM model (see Table~\ref{tab:stats}). In terms of $p$ values this translates to 3\%, 18\%, and 27\% for consistency with $\Lambda$CDM in the (C), (A), and $\gamma w \xi$ scenarios, respectively, where for all other models this number is greater than $50\%$.
However, several best-fit parameter values of the $\gamma w [\Sigma] \xi [{\rm k}]$, (A), and (C) models lie beyond their corresponding 1D-marginalized 68\% confidence limits. In the case of, e.g., the $\gamma w \xi$ model with parameter space $\mathcal{C}\cup\{w_0,w_a,\xi_0\}$, the best fit of $\xi_0$ lies even beyond the 1D-marginalized 99\% confidence level, indicating that the best-fit model
occupies only a very small subspace of the parameter space.

Remarkably, this increase in maximum likelihood ($\Delta\chi^2_{\rm max}\lesssim-2$) seems to appear only in cases where we allow a free, positive, $\xi_0$, phantom crossing in $w_{\rm DE}$, and $w_0\approx-w_a$, i.e., an effective dark energy equation of state that drives toward a matterlike equation of state at early times. We further observe that in these scenarios, there is no signature of $n_s\neq1$. Quite to the contrary, the overall best fit obtained in model (C), i.e., $n_s=0.996$, portrays an almost perfect scale-invariant Harrison-Zel'dovich spectrum~\cite{harrison:69, zeldovich:72}. Moreover, we observe a slight preference for smaller dark matter densities, smaller amplitudes for the primordial superhorizon power of the curvature perturbation, and larger ratios of sound horizon to angular diameter distance at recombination with respect to parameter values inferred for the concordance model. 
If we additionally assume a flat Universe and matter density perturbations that relate to lensing potentials as in $\Lambda$CDM, i.e., $\Sigma_0=0$, these scenarios induce values for $\overline{\langle \mathcal{L}\rangle_{\rm s}}$ beyond unity, suggesting an increase of the goodness of fit through the inclusion of their extra parameters.

The increase in maximum likelihood observed for models with $\xi_0>0$ and $w_a\approx-w_0>1$ is mainly attributed to slightly better fits of the CMB anisotropy data on all scales and to a smaller extent also to marginally better fits in distance measures,
as can be perceived from the figures in~\textsection\ref{sec:predictions}.
Merely the gISW data counteracts this trend in favor of $\Lambda$CDM and when removed from the joint data set for, e.g., the (C) model, here denoted by (C)$^*$, we obtain the most extreme values for our statistics, Eqs.~(\ref{eq:goodness_a}) through (\ref{eq:goodness_b}) (see Table~\ref{tab:stats}), while $\Lambda$CDM parameter values remain consistent at the 95\% confidence level (cf. Table~\ref{tab:results_three}).

\begin{figure*}
 \resizebox{\hsize}{!}{\includegraphics{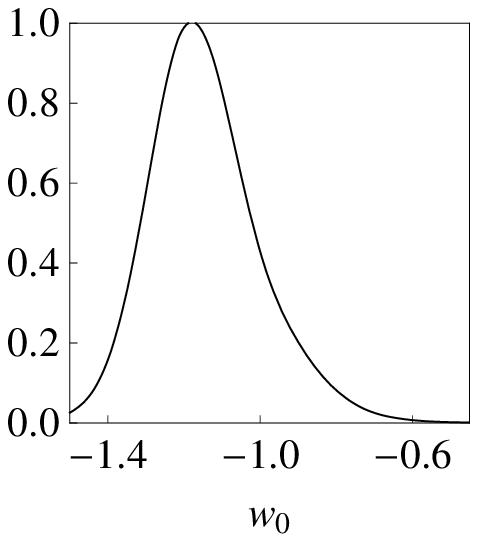}\includegraphics{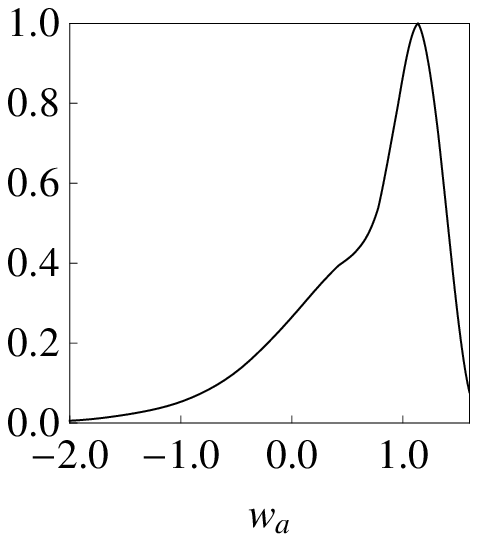}\includegraphics{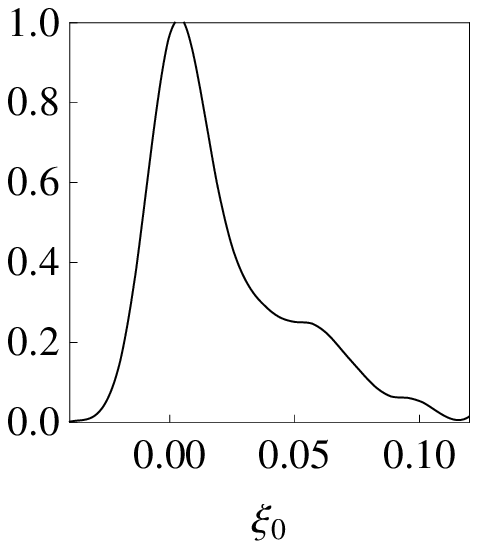}\includegraphics{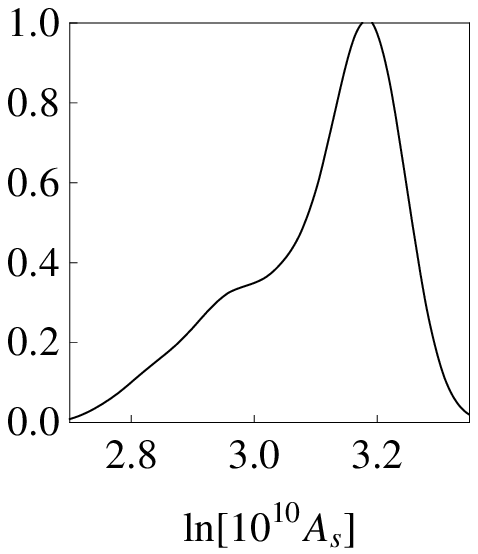}\includegraphics{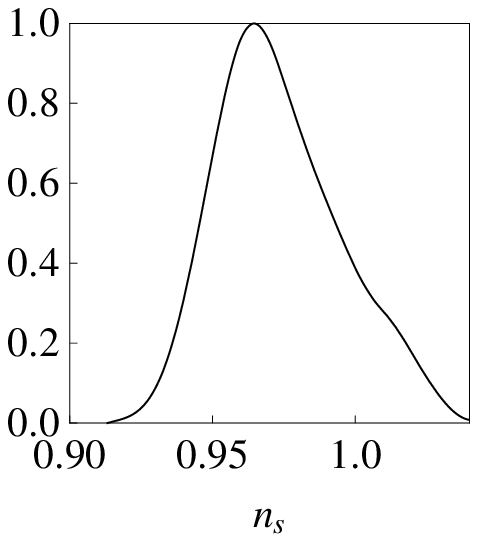}}
 \caption{Marginalized likelihoods for different parameters of the $\gamma w \xi$ model that exhibit skew distributions.}
 \label{fig:likes}
\end{figure*}

\begin{figure*}
 \resizebox{\hsize}{!}{\includegraphics{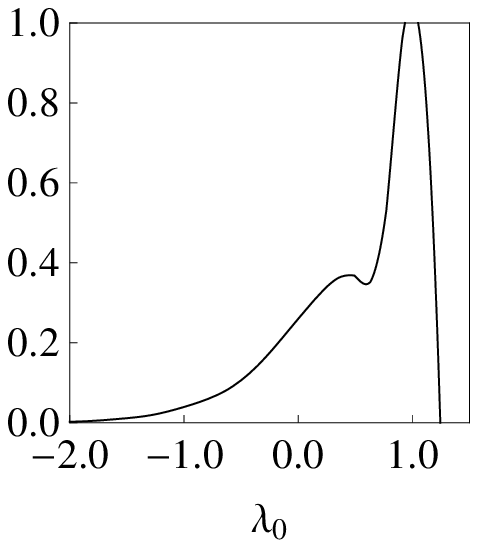}\includegraphics{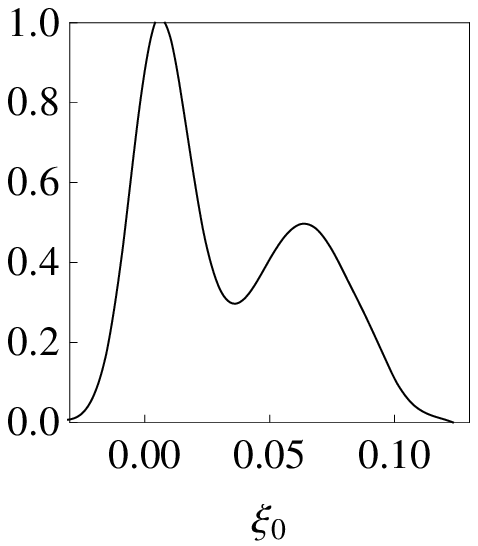}\includegraphics{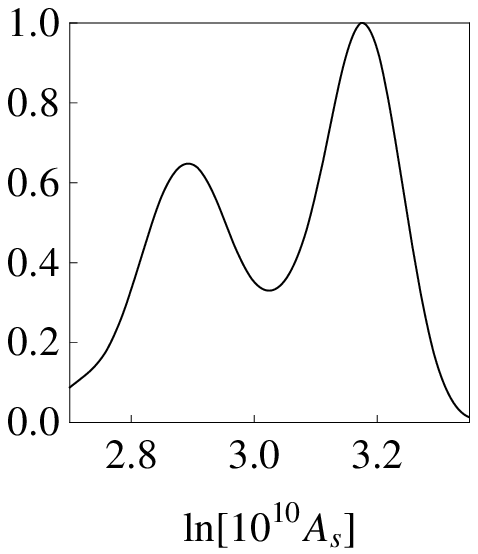}\includegraphics{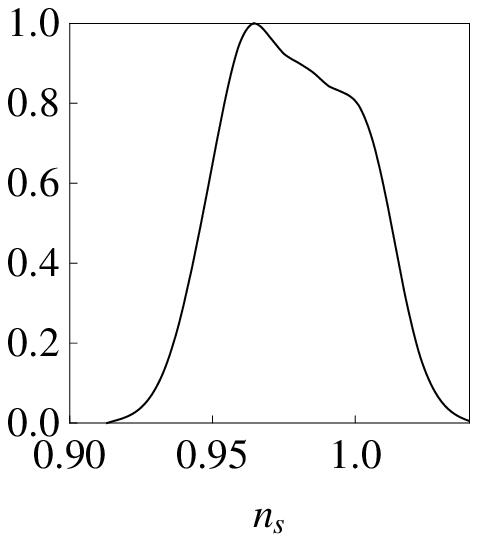}\includegraphics{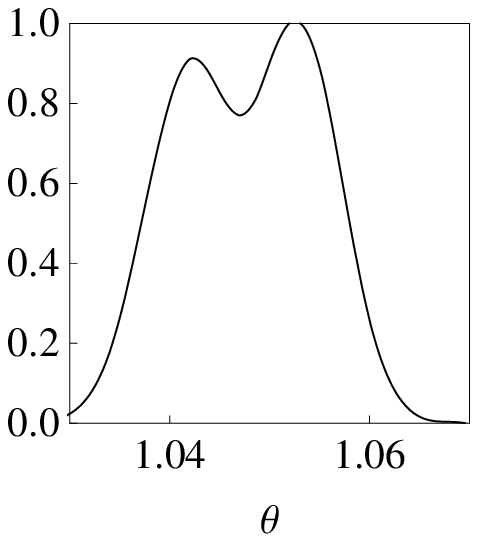}}
 \caption{The same as Fig.~\ref{fig:likes}, but for the (C) model. The posterior distribution is bimodal.}
 \label{fig:likes_C}
\end{figure*}

\begin{figure*}
 \resizebox{\hsize}{!}{\includegraphics{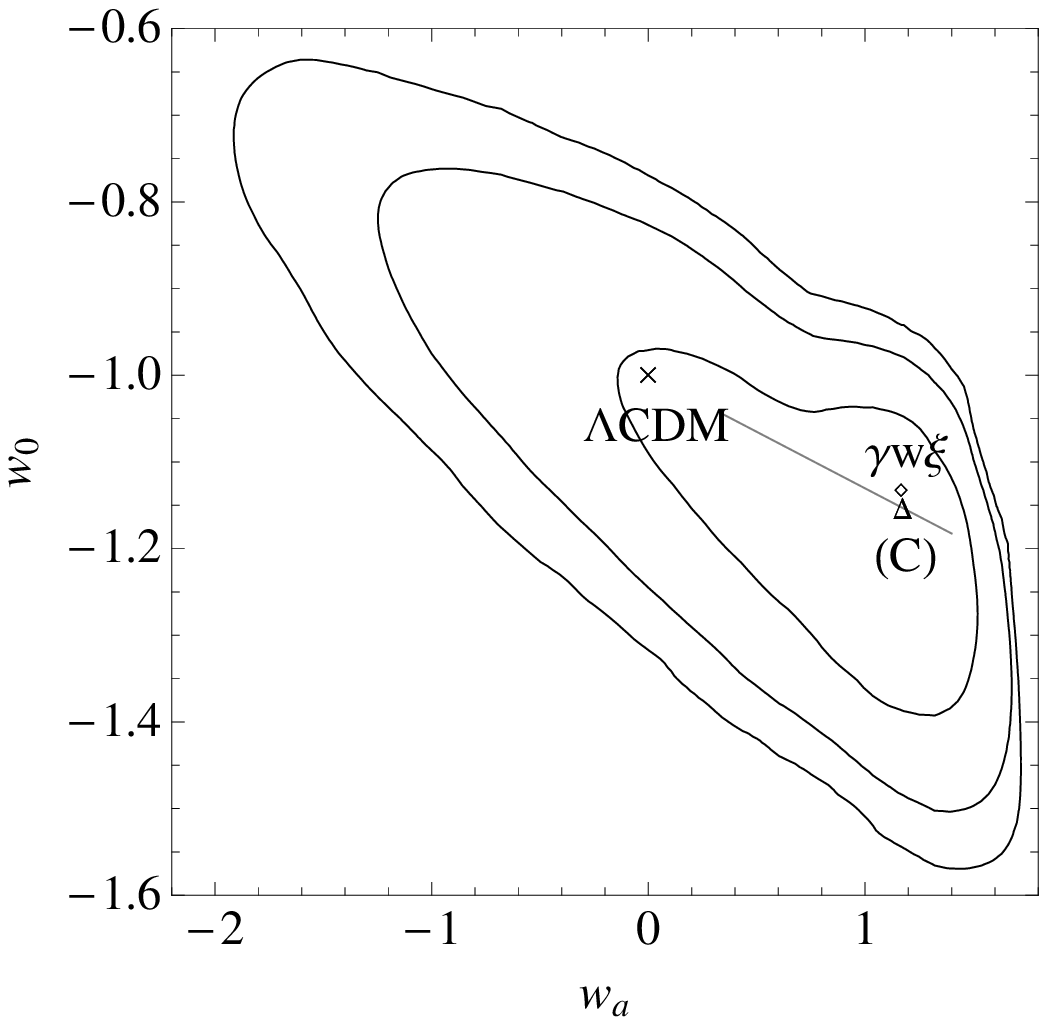}\resizebox{0.57\hsize}{!}{\includegraphics{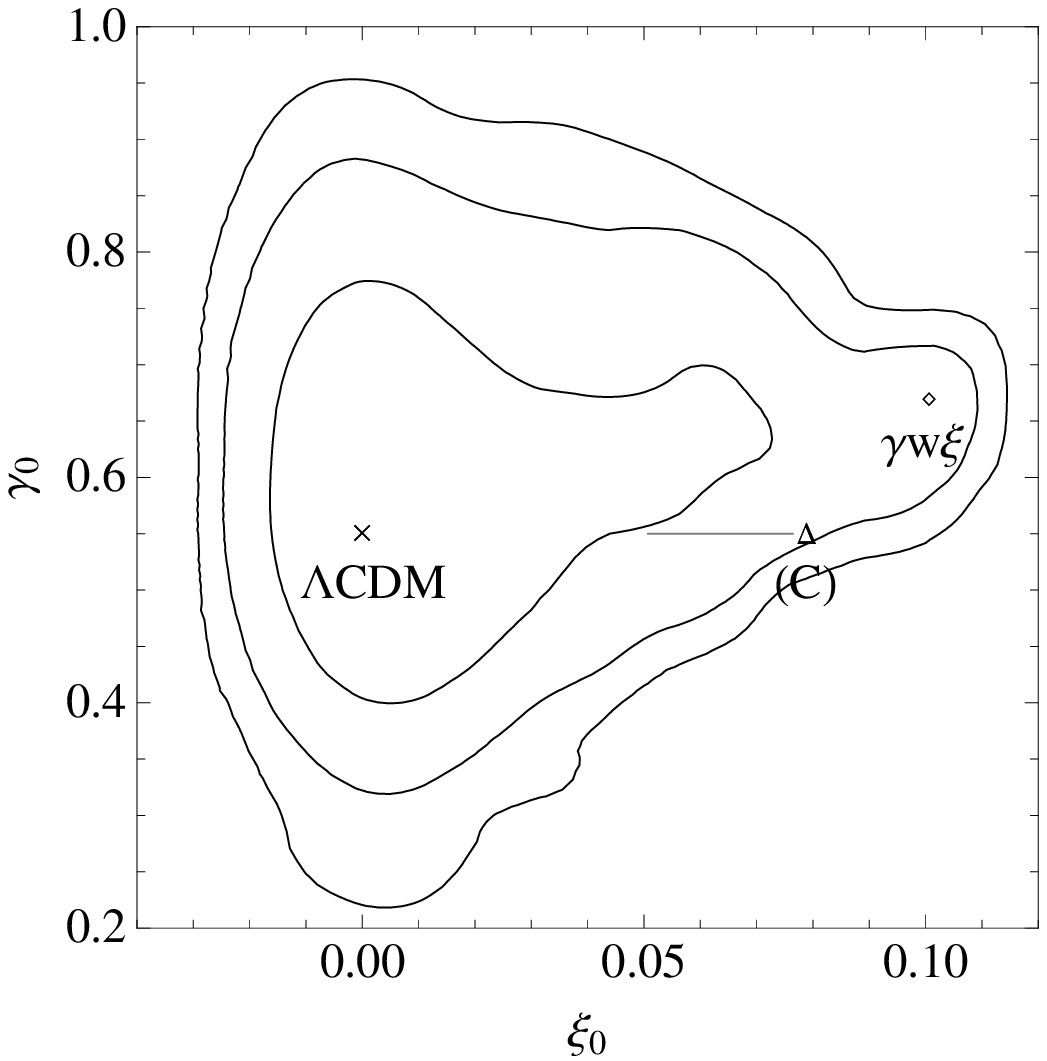}}\includegraphics{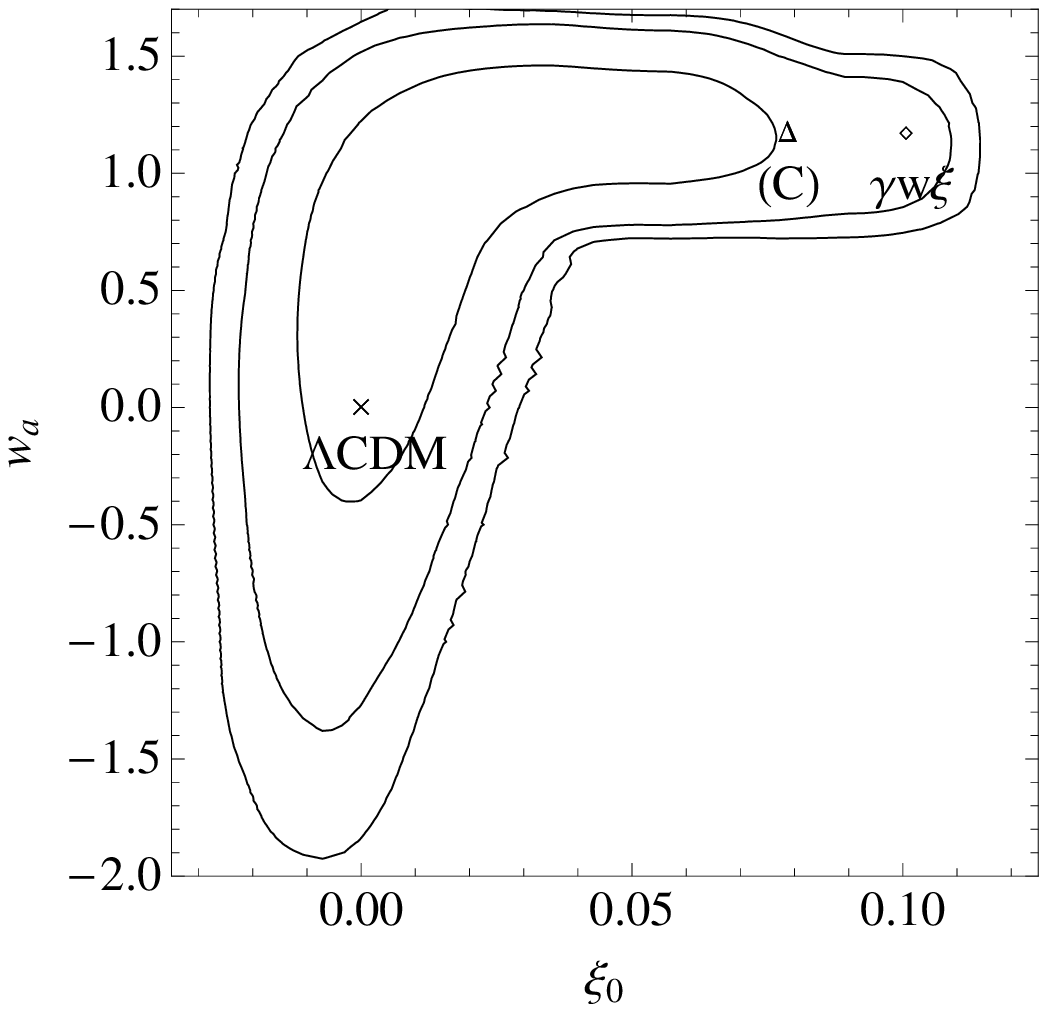}}
 \caption{2D-marginalized contours for the background parameters $(w_a,w_0)$ (left panel), the growth rate parameters $(\xi_0,\gamma_0)$ (middle panel), and $(\xi_0,w_a)$ (right panel), respectively. The boundaries indicate the 68\%, 95\%, and 99\% confidence limits, respectively. Also shown are the best-fit values for the $\Lambda$CDM, $\gamma w \xi$, and (C) models. The solid gray line indicates the 1D-marginalized 68\% minimal credible interval for $\lambda_0$ (left panel) and $\xi_0$ (middle panel), respectively.}
 \label{fig:contours}
\end{figure*}

\begin{figure*}
 \resizebox{\hsize}{!}{\includegraphics{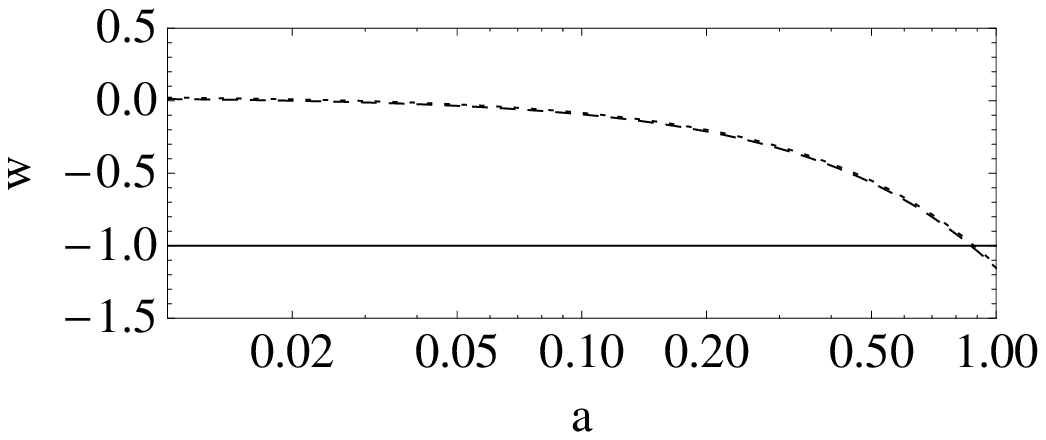}\includegraphics{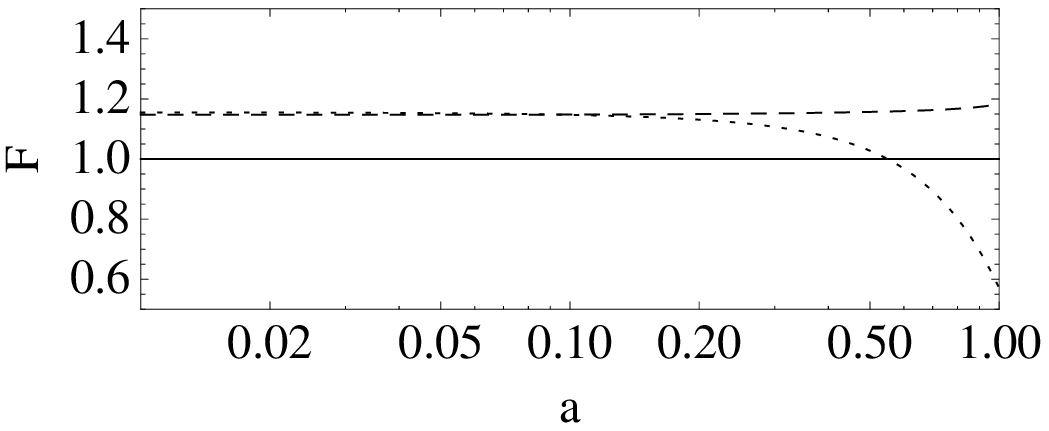}\includegraphics{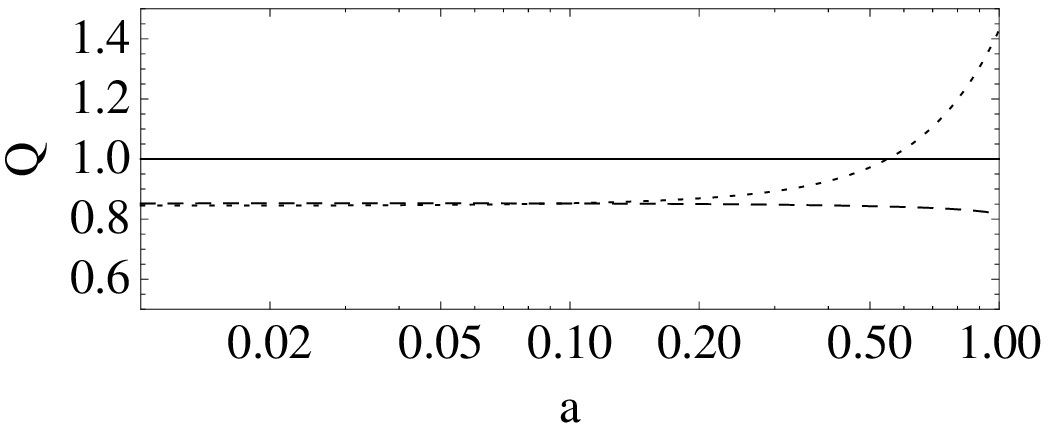}}
 \resizebox{\hsize}{!}{\includegraphics{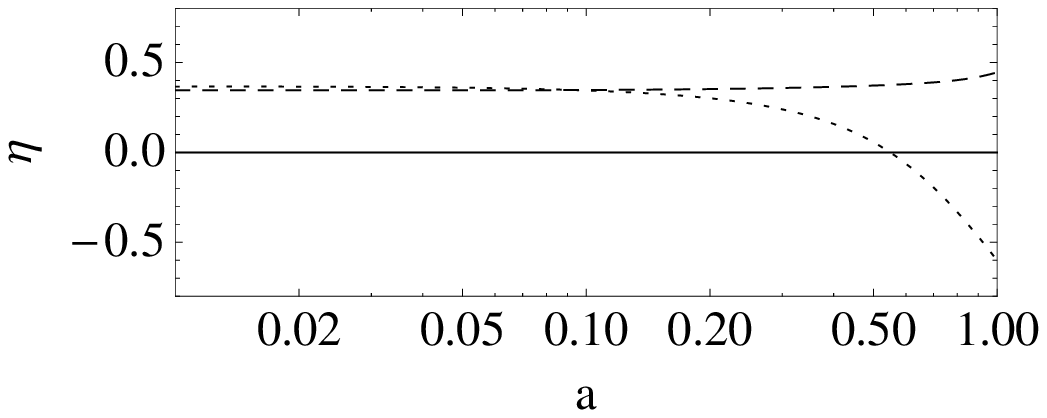}\includegraphics{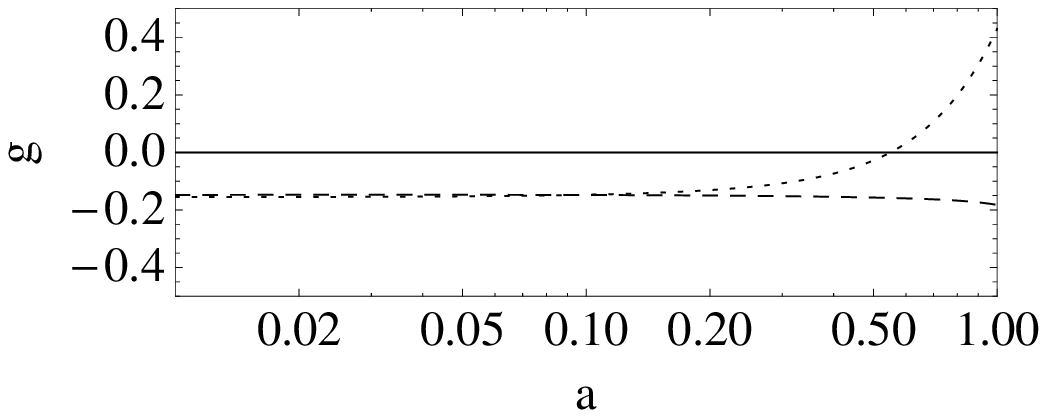}\includegraphics{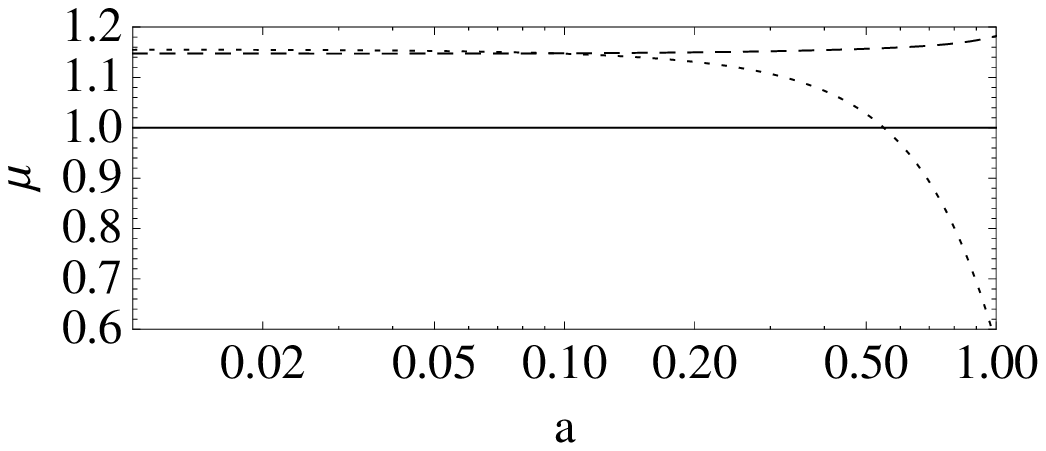}}
 \caption{Best-fit predictions for the parameters that quantify departures from the concordance model (see~\textsection\ref{sec:theory}) in the $\gamma w \xi$ (dotted lines) and (C) (dashed lines) models. Solid lines indicate the $\Lambda$CDM predictions. We also show predictions for the metric ratio $g=-\eta/(\eta+2)$~\cite{hu:07b} and the parameter $\mu=Q(1+\eta)$~\cite{song:10}. Note that $w_{\rm DE}(a)$ approaches a matterlike equation of state at early times. Predictions for $w_{\rm DE}(a)$ from the $\gamma w \xi$ and (C) models overlap.}
 \label{fig:par}
\end{figure*}

Figures~\ref{fig:likes} and \ref{fig:likes_C} show the marginalized likelihoods for the extra parameters and the parameters that exhibit a distinctive skewness or multiple modes in the case of the $\gamma w \xi$ and (C) model, respectively.
In Fig.~\ref{fig:contours}, we plot 2D-marginalized contours of the extra parameters $w_0$, $w_a$, $\gamma_0$, and $\xi_0$ within the $\gamma w \xi$ model. We also indicate the 1D-marginalized minimal credible intervals on $\lambda_0$ and $\xi_0$ obtained in model (C). Note that the best-fit points lie outside the 2D-marginalized 68\% confidence level contours of $\xi_0$ and $\gamma_0$, as well as of $\xi_0$ and $w_a$, which is due to the fact that the best-fit model occupies only a small parameter subspace and a distinctive skewness of the posterior distribution in the parameter directions. Note that there is a strong degeneracy between the parameters $\xi_0$ and $w_a$ as we expected from Eq.~(\ref{eq:gamma_par}).
Figure~\ref{fig:par} illustrates the overall best-fit parameter values from model (C) and $\gamma w \xi$ for the modifications in the dark energy equation of state $w_{\rm DE}(a)$, the factor $\mathcal{F}$ that enters the ordinary differential equation of quasistatic matter overdensity perturbations, the scaling of the effective Newton's constant $Q$, and $\eta$ that modifies the ratio of the scalar potentials. Also shown are the standard values for these quantities.

We find consistency of the $\Lambda$CDM values at the 95\% confidence level for all extra parameters in all scenarios. We, therefore, do not claim detection of nonstandard cosmological effects or the necessity of extra degrees of freedom based on the increases in the maximum and mean likelihoods (see Table~\ref{tab:stats}). This would require a physically better-motivated model with well-understood priors and a more sophisticated analysis of the goodness of fit with, e.g., efficient exploration of the different modes and determination of the Bayes factor for model comparison (see, e.g.,~\cite{feroz:07, feroz:08, mukherjee:05, shaw:07}). Note that even for a hypothetical, theoretically well-motivated model that would naturally access the best-fit region in the parameter space, there would only be a moderate preference over the concordance model, i.e., $\Delta\chi^2_{\rm max}\simeq-7$. It is not clear if such a model exists or may be developed, but we emphasize that we may \emph{a priori} be excluding viable departures from the concordance model when constraining supplementary parameters for the background and growth of structure separately.

\section{Discussion}\label{sec:discussion}

We allowed departures from the concordance model by introducing six additional degrees of freedom parametrizing a time-dependent effective dark energy equation of state, modifications of the Poisson equation for the lensing potential, and modifications of the growth of linear matter density perturbations, as well as spatial curvature. We constrained alternate combinations of these parameters by performing MCMC likelihood analyses on cosmological data amenable to linear perturbation theory. In particular, we utilized all of the CMB data, including the lowest multipoles, its correlation with galaxies and the comparison of weak lensing to large-scale velocities in addition to geometrical probes from supernovae and BAO distances, as well as from the Hubble constant.

We find consistency of the concordance model at the 95\% confidence level. For specific combinations of the supplementary free parameters, we, however, find an increase in the maximum likelihood of up to $\Delta\chi^2_{\rm max}\simeq-7$ over the maximum-likelihood $\Lambda$CDM model in the joint data. We observe that this increase in maximum likelihood only appears in cases where we allow a free, positive, $\xi_0$, phantom crossing in $w_{\rm DE}(a)$, and $w_0\approx-w_a$, i.e., an effective dark energy equation of state that drives toward a matterlike equation of state at early times. Moreover, we find that in these scenarios there is no preference for $n_s\neq1$. For scenarios where we adopt the assumptions of a flat Universe and a standard relation of matter density perturbations to lensing potentials, both maximum and mean likelihoods are greater than in $\Lambda$CDM, with a maximal ratio of mean likelihoods of the samples of $\overline{\langle \mathcal{L}\rangle_{\rm s}} \simeq 6$, suggesting that the introduction of their extra parameters might be justified. The increase in maximum likelihood can be attributed to better fits of the CMB anisotropy data on all scales and to a smaller extent also to better fits in distance measures. We therefore expect future CMB data from, e.g., the Planck mission~\cite{PLANCK:10} to yield more decisive constraints on our modifications (cf., e.g.,~\cite{mortonson:08}). The gISW data sets counteract this trend in favor of the $\Lambda$CDM model and when removed, we obtain $\Delta\chi^2_{\rm max}\simeq-7$ and $\overline{\langle \mathcal{L}\rangle_{\rm s}} \simeq 11$ for the (C) model.

Given the consistency of $\Lambda$CDM parameter values with their marginalized constraints, also under removal of gISW data, and the lack of a better-motivated theory and priors, we do, however, not claim detection of nonstandard cosmological effects or the necessity of the introduction of extra degrees of freedom, but we emphasize that when constraining the background parameters and growth parameters separately, we may \emph{a priori} be excluding viable departures from $\Lambda$CDM. We leave the search for a well-motivated model that matches the above requirements and the analysis of its performance in confrontation with new data, as well as the study of effects from scale-dependent deviations from the concordance model on our results to future work. We also point out that more sophisticated sampling methods may significantly improve our statistical analysis and offer more information through efficient exploration of the different modes and the determination of the Bayes factor for model comparison.

\section*{Acknowledgments}

I thank Uro\v{s} Seljak, An\v{z}e Slosar, Tristan Smith, Eric Linder, David Rapetti, Scott Daniel, Dragan Huterer, and Gregory Martinez for useful discussions and comments. Computational resources were provided on the Schr\"odinger and zBox2 supercomputers at the University of Z\"urich. This work was supported by the Swiss National Foundation under Contract No. 2000\_124835/1.

\appendix

\section{Connecting our parametrization to the linear PPF framework}\label{sec:ppf_connection}

Given the expansion history, the PPF framework~\cite{hu:07b, hu:08} is defined by three functions and one parameter. From these quantities, the dynamics are determined by conservation of energy and momentum and the Bianchi identities. The defining quantities are $g(a,k)$, which quantifies the effective anisotropic stress of the modifications and distinguishes the two gravitational potentials, $f_{\zeta}(a)$, which describes the relationship between the matter and the metric on superhorizon scales, and $f_G(a)$, which defines it in the linearized Newtonian regime. The additional parameter $c_{\Gamma}$ is the transition scale between the superhorizon and Newtonian behaviors.

From the relations defined in~\textsection\ref{sec:theory}, we infer $g = -\eta/(\eta+2)=1 - \mathcal{F}/\Sigma$ and $f_G = (1-\Sigma)/\Sigma$ in the quasistatic regime. At superhorizon scales, we impose $f_{\zeta} = 0$ and we set the transition at the horizon, $c_{\Gamma} = 1$. Given that $g\rightarrow 0$ sufficiently fast for $a\ll1$, if $|\xi|\ll1$ and $\Sigma\rightarrow1$, we apply the relation $g=1 - \mathcal{F}/\Sigma$ to all scales.

We implement these relations and the definitions in~\textsection\ref{sec:theory} into the PPF modified {\sc{CAMB}} code~\cite{fang:08b}. This procedure produces the correct power spectra for $\Lambda$CDM and self-accelerating DGP gravity. The latter, however, only when restricting to its subhorizon effects. In connecting our parametrization to the PPF linear theory, we can take advantage of a fully consistent framework for modifications in the gravitational dynamics. In this way, we can also prevent implicit violations of the conservation of energy and momentum and avoid gauge artifacts in our parametrization.

\vfill
\bibliographystyle{arxiv_physrev}
\bibliography{lcdm_check}

\end{document}